
\input psfig
\input mn


\newif\ifprintcomments
 
 
\printcommentstrue


\def\etal{et al.~}


\def\Msun{{\rm\,M_\odot}}



\def\deg{^{\circ}}


\def\spose#1{\hbox to 0pt{#1\hss}}
\def\lta{\mathrel{\spose{\lower 3pt\hbox{$\sim$}}
    \raise 2.0pt\hbox{$<$}}}
\def\gta{\mathrel{\spose{\lower 3pt\hbox{$\sim$}}
    \raise 2.0pt\hbox{$>$}}}


\newdimen\hssize
\hssize=8.4truecm
\newdimen\hdsize
\hdsize=17.7truecm


\def\today{\ifcase\month\or
 January\or February\or March\or April\or May\or June\or
 July\or August\or September\or October\or November\or December\fi
 \space\number\day, \number\year}
 

\newcount\eqnumber
\eqnumber=1
\def\chaphead{} 

\def\new{\hbox{(\rm\chaphead\the\eqnumber)}\global\advance\eqnumber by 1}
 
\def\first{\hbox{(\rm\chaphead\the\eqnumber a)}\global\advance\eqnumber by 1}
\def\last#1{\advance\eqnumber by -1 
            \hbox{(\rm\chaphead\the\eqnumber#1)}\advance
     \eqnumber by 1}
 
\def\ref#1{\advance\eqnumber by -#1 \chaphead\the\eqnumber
     \advance\eqnumber by #1}
\def\nref#1{\advance\eqnumber by -#1 \chaphead\the\eqnumber
     \advance\eqnumber by #1}

\def\eqnam#1{\xdef#1{\chaphead\the\eqnumber}}

 
 
 




\pageoffset{-0.85truecm}{0.45truecm}



\pagerange{}
\pubyear{version: \today; {\tt DO NOT DISTRIBUTE}}
\volume{}

\begintopmatter

\title{On the uncertainties of the central density in axisymmetric galaxies
due to deprojection}

\author{Frank C. van den Bosch}

\affiliation{Sterrewacht Leiden, Postbus 9513, 2300 RA Leiden, 
             The Netherlands}
\vskip 0.1truecm
\affiliation{vdbosch@strw.LeidenUniv.nl}

\shortauthor{F.C. van den Bosch}

\shorttitle{%
Uncertainties of the central density in axisymmetric galaxies}


\abstract{%
The deprojection of the surface brightness of axisymmetric galaxies is 
indeterminate unless the galaxy is seen edge-on. In practice, this problem 
is often circumvented by making {\it ad hoc} assumptions about the density 
distribution. However, one can redistribute the density and still project 
to the same surface brightness. This is similar to adding so-called konus 
densities to the assumed density distribution.

In this paper we investigate the maximum konus density that one can 
add to elliptical galaxies. In particular we focus on the uncertainties
in the central densities of axisymmetric, elliptical galaxies due to the 
non-uniqueness of the deprojection. 

For St\"ackel potentials a sufficient condition for positivity of the
phase space distribution function exists, which is used as criterion to 
determine the maximum konus density that one can add to a perfect oblate 
spheroid. For small inclination angles we find an uncertainty in the central 
density of up to a factor two. 

Elliptical galaxies in general have a central density cusp. We therefore
also investigate the maximum konus densities of cusped ellipticals.
For these models we use an approximate criterion, which we have tested
on the perfect oblate spheroid models. For sufficiently small scalelengths,
the central density of the maximum konus density that can
be added to a cusped elliptical is very high. In order to
estimate the dynamical influence of konus densities on the central region, 
we calculate the mass fraction $M_{\rm kon}/M_{\rm gal}$ they can add to the 
center. We show that this mass fraction is at most a few percent.
We also investigate the dynamical effect of cusped konus densities that
have $\rho(r) \propto r^{-\alpha}$ at small radii. We show that konus 
densities can only be moderately cusped ($\alpha < 1$), and that an 
increase in cusp slope $\alpha$, results in a {\it decrease} of the mass
fraction added to the center by the konus density.

We illustrate all this by the specific example of M32, and show that the 
uncertainty in the central mass due to deprojection is negligible compared 
to the inferred mass of the central black hole (BH) in this galaxy. 
}

\keywords{%
Galaxies: fundamental parameters -- galaxies: ellipticals -- 
galaxies: nuclei -- galaxies: photometry -- galaxies: kinematics and dynamics
}

\maketitle  


\eqnumber=1
\def\chaphead{\hbox{1.}}

\section{1 Introduction}

\tx In recent years considerable progress has been made in the construction
of axisymmetric dynamical models of galaxies (e.g., de Zeeuw 1994).
Comparison of such models with observed kinematics has resulted in strong 
indications for the presence of a nuclear black hole (BH) in a number of 
galaxies (e.g., Kormendy \& Richstone 1995). 
In order to construct such models one needs to recover the density
distribution from the observed surface brightness distribution.
However, Rybicki (1986) showed, by use of the Fourier Slice Theorem, that 
the deprojection of axisymmetric bodies is indeterminate unless the 
inclination angle $i = 90\deg$ (i.e., the galaxy is seen edge-on).
In practice this problem is often circumvented by making an {\it ad hoc}
assumption about the density distribution, e.g., considering only densities
stratified on similar concentric ellipsoids. However, the Fourier Slice 
Theorem shows that any density distribution whose Fourier Transform is only 
non-zero inside a cone with half-opening angle $90\deg - i$ and aligned with 
the symmetry axis of the Fourier Transform of the density distribution (the 
so called `cone of ignorance'), projects to zero surface brightness. 
Following the nomenclature of Gerhard \& Binney (1996, hereafter GB) we will 
call such densities `konus densities'. Franx (1988) gave an example of such 
a konus density which, when added to a triaxial St\"ackel model, projects 
to zero surface brightness. GB investigated the effect of adding or subtracting
a specific family of konus densities $\rho_{\rm k}(R,z|i_0)$ to a density
distribution stratified on similar concentric ellipsoids $\rho_{0}(R,z)$. 
Although all densities 
\eqnam\totdens
$$\rho(R,z) = \rho_{0}(R,z) + f\; \rho_{\rm k}(R,z|i_0),\eqno\new$$ 
project to the same surface brightness
if $i \leq i_0$, the underlying density distributions (and therefore the
corresponding dynamics) can differ considerably.

In this paper we investigate to what extent one can add or subtract konus
densities. In particular we focus on the uncertainty of the central
density of axisymmetric galaxies due to the non-uniqueness of the 
deprojection. The kinematical evidence for the presence of nuclear
BHs arises from the finding that the central density is too low in order 
to explain the observed kinematics. This density distribution is
generally derived from the deprojection of the observed surface brightness,
which requires assumptions on the inclination angle and the density
distribution itself (i.e., the density is stratified on similar concentric 
ellipsoids). We apply our analysis to M32, in which the presence of a
BH has been claimed. The mass of the BH that is inferred is of the order
of 20\% of the total mass inside $1''$. We investigate whether
adding konus densities can redistribute the central mass such that we can
elude the requirement of a BH in M32. We show that, although the maximum
konus density that one can add can have very high central density (the konus 
density can even be cusped), the actual mass it adds to the nucleus (and 
therefore its dynamical influence) is negligible compared to the inferred BH.

This paper is organized as follows.
In Section 2, we describe the criterion we use to determine the maximum
konus density one can add to a certain galaxy model. Section 3 deals with 
the specific konus densities we use for our investigation. In Section 4 we 
discuss the uncertainties of the central density due to the freedom one has 
in redistributing the density distribution. In Section 5 an approximate 
criterion is discussed, which is applied to cusped galaxy models in Section 6. 
We discuss the application of all this to the specific example of M32 in 
Section 7. Finally, in Section 8, we summarize and discuss our results.
%
\beginfigure{1}
\centerline{\psfig{figure=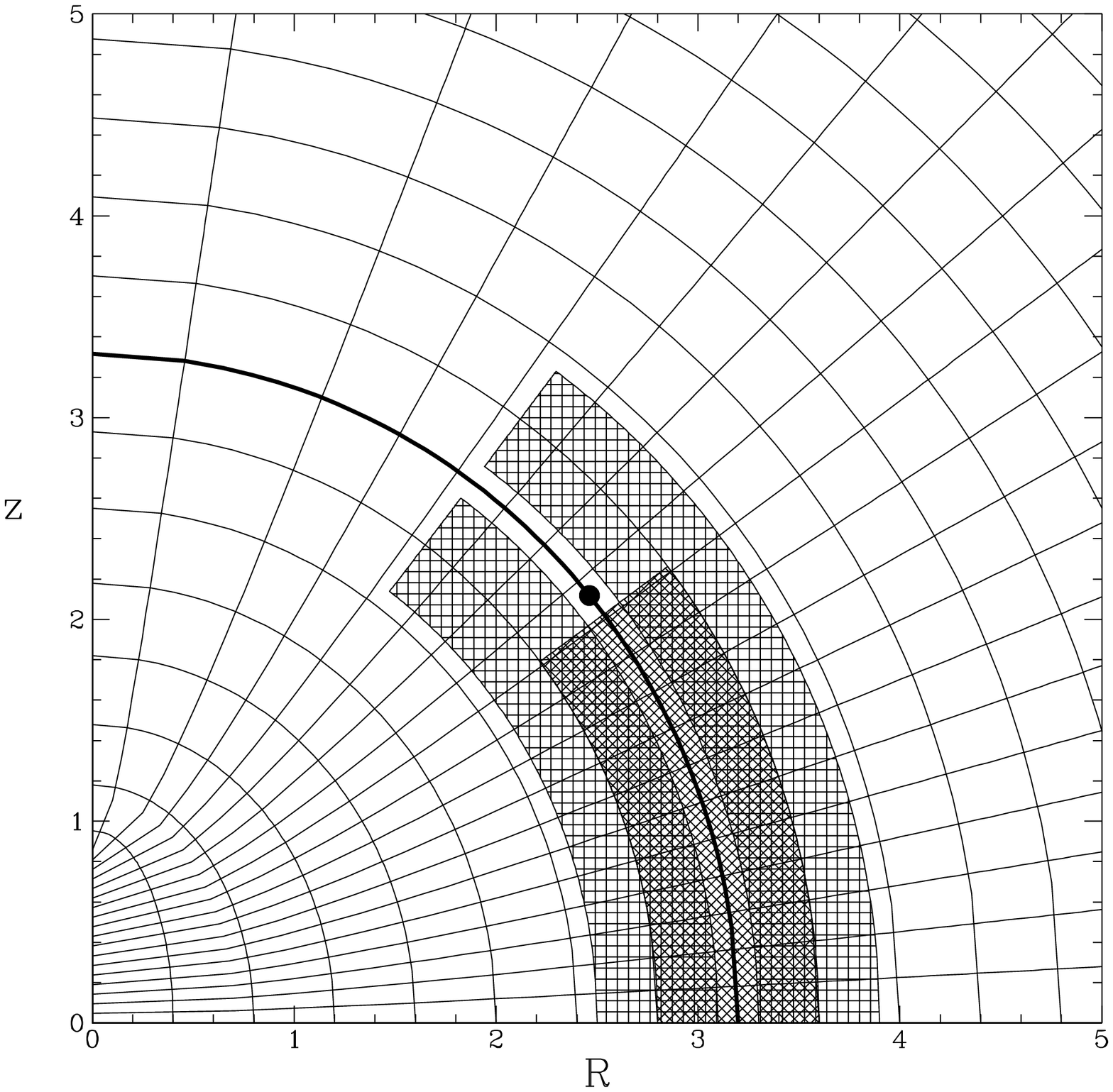,width=\hssize}}\smallskip
\caption{{\bf Figure 1.} The grid are curves of constant $\zeta$ and $\phi$, 
which are the curves in between which orbits in an axisymmetric potential
are enclosed. The dashed regions are three such orbits. The thick, solid
curve is the curve with $\zeta = \zeta_0$ and the solid dot is the point
$(\zeta_0,\phi_0)$ (see text). As can be seen, having zero density in
$(\zeta_0,\phi_0)$ results in zero density along the entire curve
$\zeta = \zeta_0$ with $\phi > \phi_0$. }
\endfigure
%
%

\eqnumber=1
\def\chaphead{\hbox{2.}}

\section{2 The criterion}

\tx Since konus densities have regions of both positive and negative
density there is a $f_{\rm max}$ such that for $f>f_{\rm max}$ their will
be regions where the total density $\rho = \rho_{\rm gal} + f\; \rho_{\rm kon}$
is negative. Therefore, $f_{\rm max}$ expresses the maximum 
konus density under the physical criterion that $\rho > 0$. 
However, this criterion ignores whether or not such a galaxy can actually be 
built from its orbit building blocks. In general, for $f=f_{\rm max}$ there
will be points $(R_0,\vert z_0 \vert)$ in the meridional plane with 
zero density. It is easily seen that such models can not be physical. 
Orbits in an axisymmetric potential densely fill a limited area inside
their zero-velocity curve. For thin tube orbits, this area shrinks to a 
one-dimensional curve. Let those curves be described by $\zeta(R,z)$
and parameterized by $\phi$, such that 
$\phi = a$ in the equatorial plane, and $\phi = b>a$ along the symmetry axis 
$R=0$. It is easily seen that it is impossible to construct a model that has 
only one point in the first quadrant of the meridional plane 
$(R_0,z_0) \equiv (\zeta_0,\phi_0)$ with zero density, and $\rho > 0$
everywhere else. This is illustrated graphically in Figure 1, where
three orbits are plotted in the meridional $(R,z)$-plane. The thick, solid 
curve is the curve with $\zeta = \zeta_0$, and the solid dot is the point 
$(\zeta_0,\phi_0)$. As can be seen, all three plotted orbits are just allowed
since they `avoid' the solid dot, and therefore do not contribute density to 
$(\zeta_0,\phi_0)$. Each orbit that contributes density to a point 
$(\zeta_0,\phi')$, with $\phi' > \phi_0$ also contributes density to all 
points $(\zeta_0, \phi < \phi')$. As a consequence, the only way to 
distribute orbits so that $\rho(\zeta_0,\phi_0) = 0$ is by having zero density
along the entire curve with $(\zeta_0,\phi > \phi_0)$. 

In general this criterion is difficult to make quantitative. For St\"ackel 
potentials, however, the third integral is known analytically and 
$(\zeta,\phi)$ are the prolate spheroidal coordinates $(\lambda,\nu)$ 
(see Appendix A). 
Bishop (1987) formulated the principle outlined above by proving that
a sufficient condition for the distribution function $f(E,L_z,I_3)$ of a 
St\"ackel model to be non-negative and hence to correspond to a proper 
equilibrium model is that the density should monotonically decrease along
lines of constant $\lambda$; i.e., 
\eqnam\criter
$${\partial\rho(\lambda,\nu) \over \partial\nu} < 0.\eqno\new$$
In principle, one can still build physical models (i.e., $f(E,L_z,I_3) > 0$) 
that do not obey this criterion (\criter), since it ignores the fact that 
orbits do
not distribute density uniformly over the area in the $(\lambda,\nu)$
plane occupied by that orbit. The criterion, although sufficient is therefore
not strict, and will lead to an underestimate of the maximum 
konus density that can be added to a St\"ackel potential.

As a specific example of a St\"ackel model we will focus on the
perfect oblate spheroids (see Appendix A), and use criterion (\criter) 
to investigate the maximum konus density one can add to such a density 
distribution.
We note that if one adds a konus density to a St\"ackel potential
it no longer is a St\"ackel potential, and therefore the orbits
will no longer be confined by curves of constant $\lambda$ and $\nu$.
However, as a first order approximation, we can still use the criterion
(\criter). As we will show, the results presented in this paper are only mildly
dependent on the actual curve along which we demand the gradient of the 
density to be negative (see Section 5).

\eqnumber=1
\def\chaphead{\hbox{3.}}

\section{3 The konus density}

\tx In the following we will concentrate on ellipticals that have
a luminous density that decays as $r^{-4}$. The surface brightness of
such systems falls of as $R^{-3}$ at large radii, which is in good 
agreement with observations. In order for the total density,
$\rho = \rho_{\rm gal} + f\; \rho_{\rm kon}$, of such systems to be 
positive, we require a konus density that falls of as $r^{-p}$ at 
large radii, where $p > 4$.

Furthermore, since we use the criterion that
\eqnam\totcrit
$$ {\partial \rho \over \partial \nu} = {\partial \rho_{\rm gal} \over 
\partial \nu} + f\; {\partial \rho_{\rm kon} \over \partial \nu} < 0,
\eqno\new$$
and since for the perfect oblate spheroid 
$$\lim_{\lambda \rightarrow \infty} 
{\partial \rho \over \partial \nu}(\lambda,\nu) \propto \lambda^{-2} = r^{-4} 
\eqno\new,$$ 
we require a konus density whose derivative with respect to the prolate
spheroidal coordinate $\nu$ decays
faster than $r^{-4}$.

The specific konus density introduced by GB (their equation 19) does not 
satisfy this criterion. 
We therefore seek another konus density. Kochanek \& Rybicki (1996; hereafter
KR) have developed a scheme to derive konus densities with a large
variety of properties through the introduction of `semi-konus densities'.
A semi-konus density is a density that is non-zero only in one of the two
halfs of the cone of ignorance; either the cone at positive or at negative
$\bf \hat z$ (here $\bf \hat z$ denotes the symmetry axis in Fourier space).
Semi-konus densities are complex functions and KR showed that the real part of
any semi-konus density is a konus density. Furthermore, semi-konus
densities have the nice property that they are closed under ordinary
multiplication: the product of two semi-konus densities is
itself a semi-konus density. In particular any power
\eqnam\highordkon
$$\rho^{n}(R,z) = \Re \bigg[\rho_{\rm sk}^{1}(R,z)\bigg]^{n},\eqno\new$$
where $\Re$ denotes the real part, is a konus density as long as 
$\rho_{\rm sk}^{1}(R,z)$ is a semi-konus density. We will refer to konus 
densities of the form (\highordkon) as konus densities of order $n$.

We have experimented with several konus densities and finally came up
with, what we will refer to as, the generalized konus density (see Appendix B).
The generalized konus density is characterized by three parameters: a 
scalelength $b$, an order $n$, and an additional free parameter $\mu$.
Increasing the latter two, increases the number of oscillations, as well
as the power $p$ of the decline at large radii. In order for the decline of 
both the density and the derivative (\totcrit) to be sufficiently fast we 
need to go to order $n \geq 6$. This is independent of $\mu$. Throughout this 
paper we will mainly focus on the generalized konus densities with $\mu = 1$. 
A contour plot of this konus density with $n=6$ in the meridional plane is 
shown in Figure 2.
%
%
\beginfigure{2}
\centerline{\psfig{figure=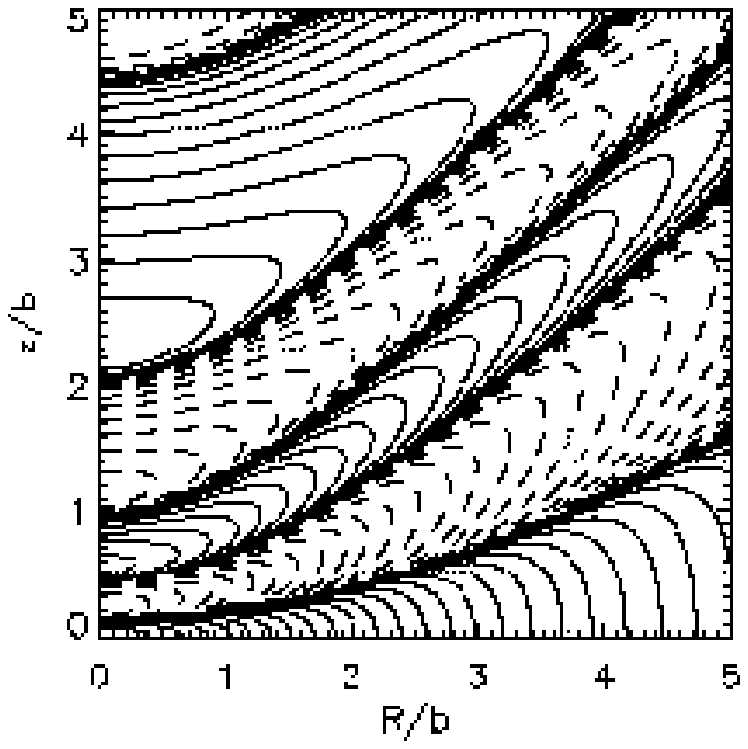,width=\hssize}}\smallskip
\caption{{\bf Figure 2.} Contours of the $6^{\rm th}$ order generalized 
konus density $\rho^{6}_{k}(R,z|45\deg)$ for $\mu=1$. This density 
distribution projects to zero surface brightness for $i \leq 45\deg$. 
Positive contours are solid, negative contours are dashed.}
\endfigure
%
%

%
%
\beginfigure*{3}
\centerline{\psfig{figure=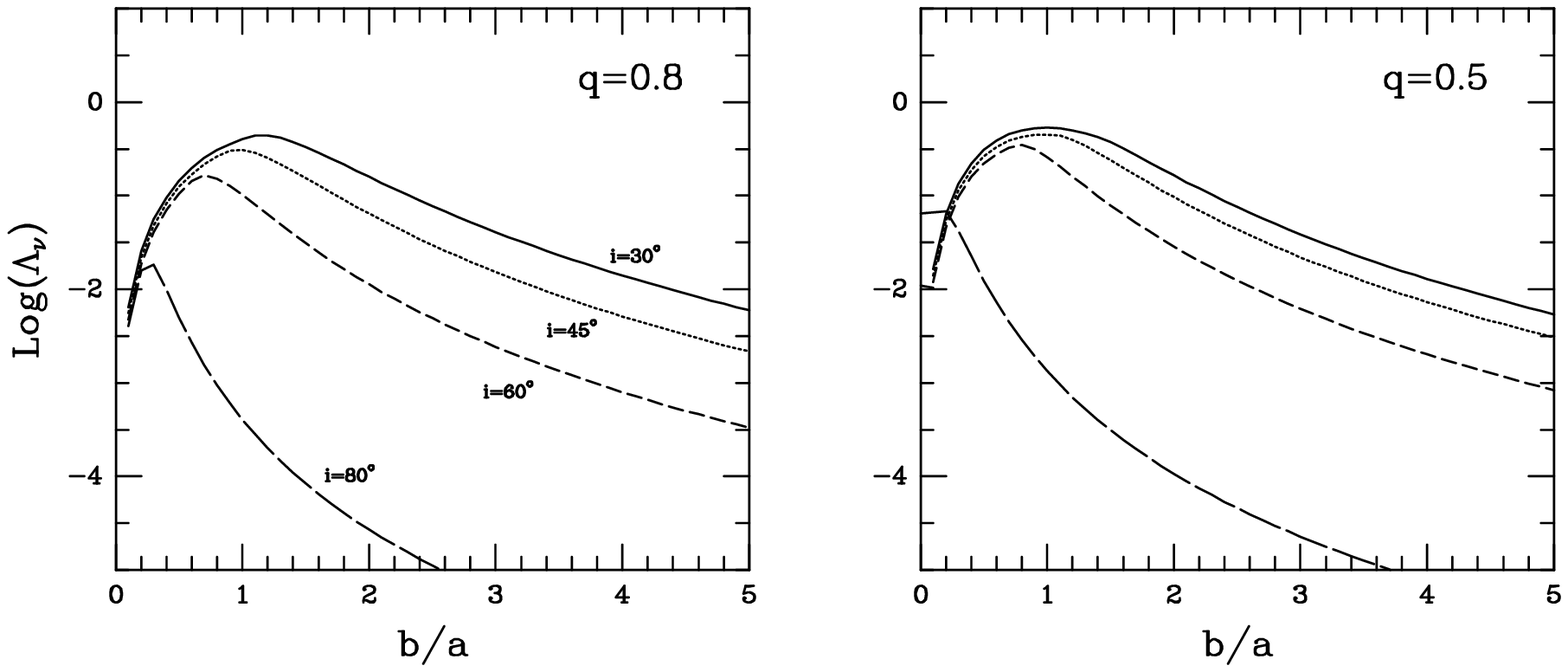,width=\hdsize}}\smallskip
\caption{{\bf Figure 3.} The logarithm of the parameter $\Lambda_{\nu}$ as 
a function of the ratio $b/a$ for the $6^{\rm th}$ order generalized konus 
density ($\mu = 1$). The results are presented for four different inclination
angles: $i=30\deg$ (solid lines), $i=45\deg$ (dotted lines),
$i=60\deg$ (short dashed lines), and $i=80\deg$ (long dashed lines).
The panel on the left is for a perfect oblate spheroid with an ellipticity
$\epsilon$ of 0.2, whereas the results in the panel on the right are
for a model with $\epsilon = 0.5$.} 
\endfigure
%
%

%
%
\beginfigure{4}
\centerline{\psfig{figure=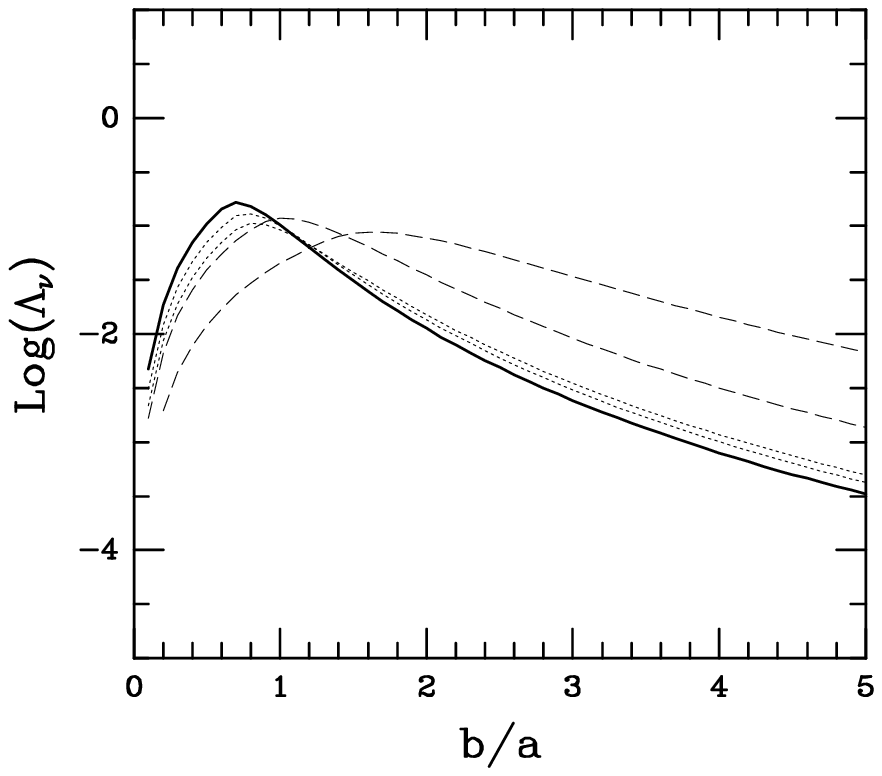,width=\hssize}}\smallskip
\caption{{\bf Figure 4.} The same as Figure 3, except that we now plot the
maximum konus densities for different orders $n$ of the generalized
konus density and for different values of $\mu$. All lines are for an oblate
spheroid with $\epsilon = 0.2$, seen under an inclination angle of $60\deg$.
The thick, solid line is for $\mu =1$ and $n=6$. The dashed lines are for
$\mu = 2$ and $\mu=4$, and the dotted lines are for higher order; $n=7$ and
$n=8$. As can be seen, central density of the maximum konus density increases
with both $\mu$ and $n$ for large scalelengths $b$. However, for small 
scalelengths ($b/a \lta 1$), $\Lambda_{\nu}$ is highest for $\mu =1$, $n=6$.}
\endfigure
%

%
%
\beginfigure*{5}
\centerline{\psfig{figure=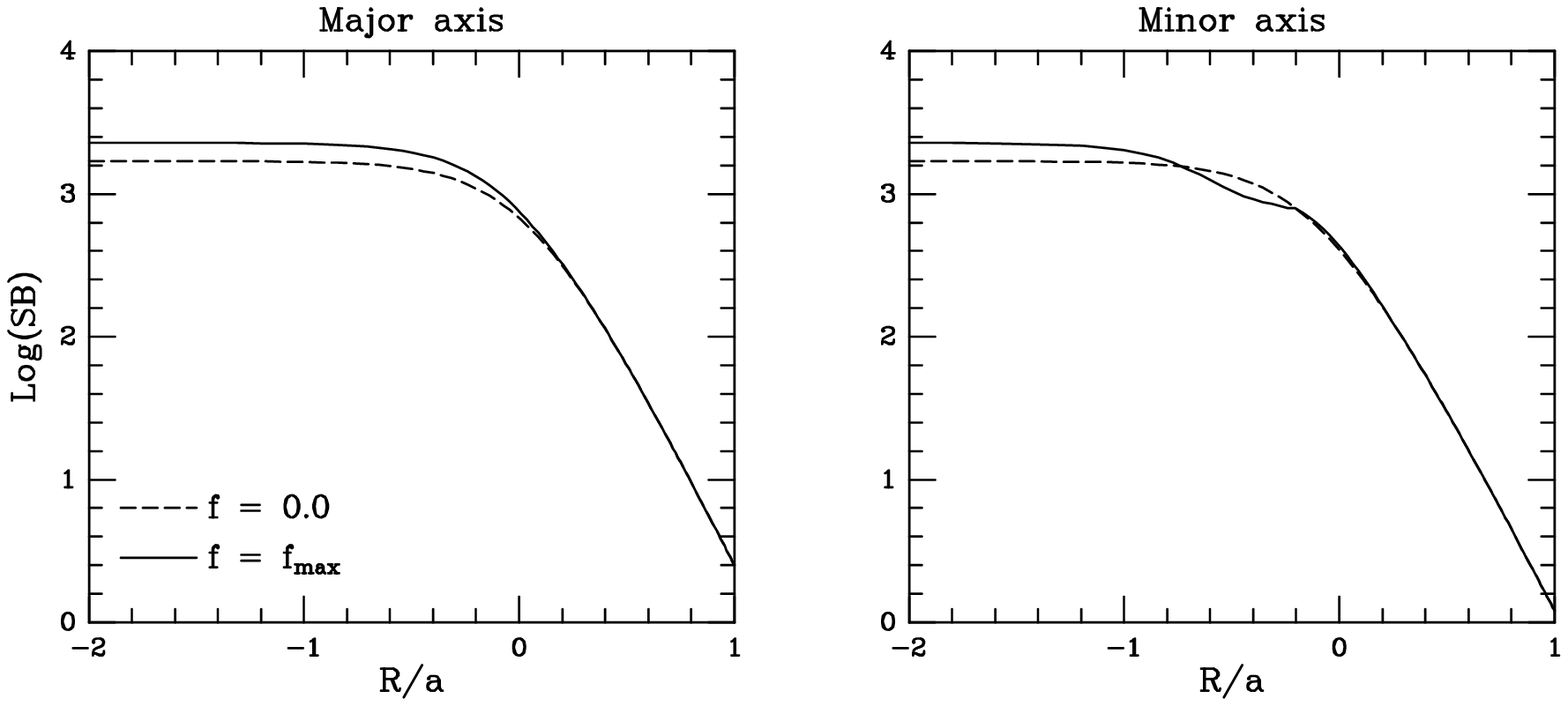,width=\hdsize}}\smallskip
\caption{{\bf Figure 5.} The luminosity profiles along the major and
the minor axis of a perfect oblate spheroid with intrinsic flattening
$q = 0.8$ seen edge-on (dashed lines). In addition, we also plot
the luminosity profile of the same spheroid but to which we
have added the maximum konus density $\rho_{\rm k}^{6}(R,z|30\deg)$.
The scalelength of the konus density $b = 1.3 a$, where $a$ is the break radius
of the spheroid. We have plotted the logarithm of the projected surface 
brightness (in arbitrary units) {\it vs.} the logarithm of the radius on the 
sky $R$ in units of $a$.}
\endfigure
%
%

\eqnumber=1
\def\chaphead{\hbox{4.}}

\section{4 Uncertainties in the central density}

We now investigate the maximum konus density one can add to
or subtract from a perfect oblate spheroid under the criterion of equation
(\criter). We therefore seek the value
\eqnam\maxf
$$f_{\rm max} = \max(f_{+},f_{-}),\eqno\new$$
where 
\eqnam\fff
$$f_{\pm} = \min_{(\lambda,\nu)} \bigg( f > 0; 
{\partial\rho_{\rm gal} \over \partial\nu} \pm f \; 
{\partial\rho_{\rm k}^{n} \over \partial\nu} = 0\bigg).\eqno\new$$
Here $\rho_{\rm k}^{n}$ is the $n^{\rm th}$ order of the generalized konus
density (see Appendix B). So $f_{+}$ describes the maximum 
konus density one can add, and  $f_{-}$ the maximum that 
can be subtracted.
They are the minimal values of $f$ for which the derivative of the total 
density with respect to $\nu$ equals zero. This value
has to be searched over the entire $(\lambda,\nu)$-space.
Once $f_{\rm max}$ is found we calculate
\eqnam\ratio
$$\Lambda_{\nu} \equiv {f_{\rm max} \; \rho_{\rm k}(0,0|i) \over 
\rho_{\rm gal}(0,0)}.\eqno\new$$
The parameter $\Lambda_{\nu}$ therefore expresses the ratio of the central
density of the maximum konus density over the central density of the 
perfect oblate spheroid.
%
%
\beginfigure*{6}
\centerline{\psfig{figure=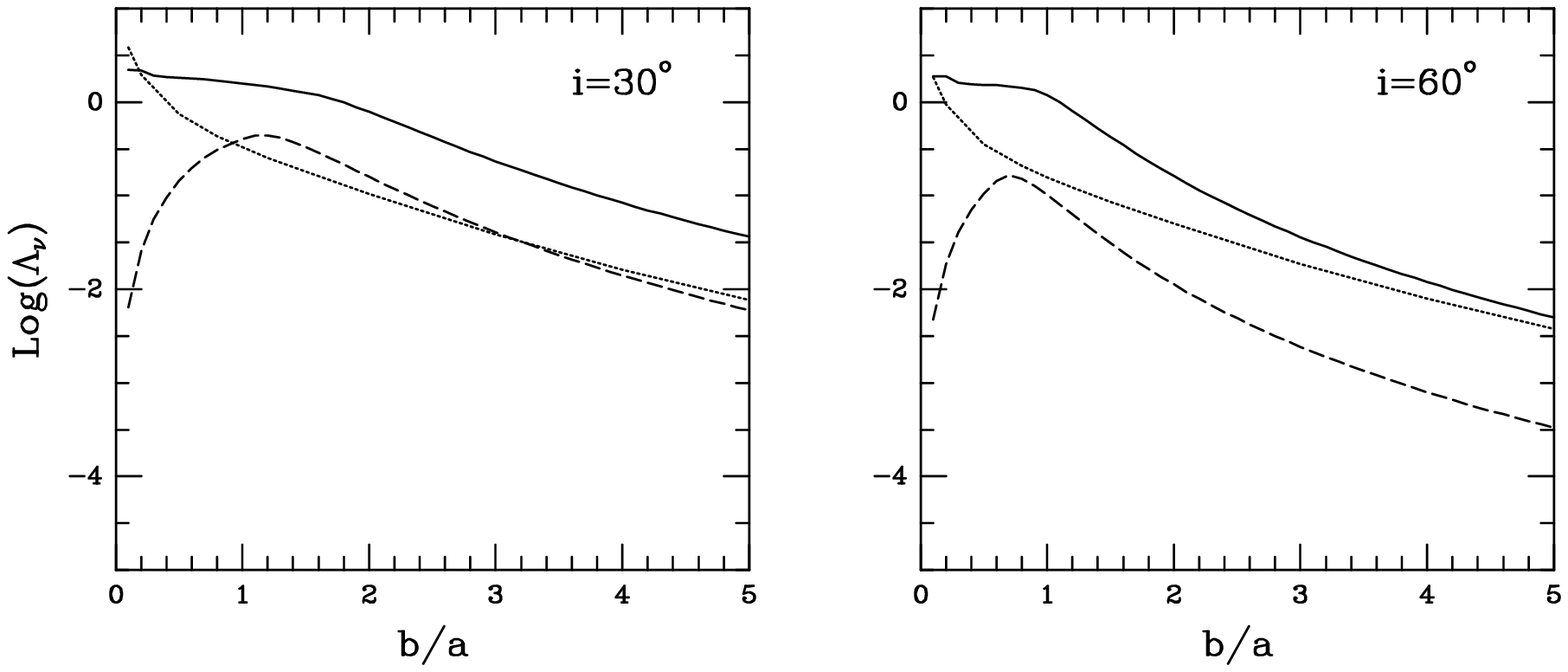,width=\hdsize}}\smallskip
\caption{{\bf Figure 6.} The maximum amount of konus density $\Lambda_{\nu}$
that can be added to a perfect oblate spheroid with $q=0.8$, as determined 
from three different criteria: $\rho > 0$ (solid lines), $\partial \rho / 
\partial \nu < 0$ (dashed lines), and the empirical criterion that the minor 
axis surface brightness profile should not have wiggles (dotted lines).
The results are shown as function of $b/a$ for two different inclination
angles ($i = 30\deg$, and $60\deg$).}
\endfigure
%
%

The results for the generalized konus density of order 6 with $\mu =1$ 
are presented in Figure 3, where we plot the logarithm of
$\Lambda_{\nu}$ as a function of the ratio $b/a$. Here $b$ is the scalelength 
of the konus density and $a = \sqrt{-\alpha}$ is the break radius
of the perfect oblate spheroid (see Appendix A). The results are shown for
four different inclination angles ($i=30\deg,45\deg,60\deg$, and
$80\deg$), and two different flattenings $q$ ($q = 0.8$ and $0.5$.)
For small inclination angles the central density of the maximum konus density 
is of the same order of magnitude as the central density of the underlying 
spheroid (i.e., $\Lambda_{\nu} \approx 1$). However, that is only true if the 
scalelength of the konus density is approximately equal to the break radius
of the perfect oblate spheroid (i.e., $b/a \approx 1$). For higher 
inclination angles, one can add less konus density to the model, reaching
$f_{\rm max} = 0$ for $i=90\deg$. We found that for $i = 80\deg$, the 
central density of the maximum konus density is no more than $\sim$ 10\% of 
$\rho_{\rm gal}(0,0)$, and only if the scalelength of the konus density 
becomes small. In Figure 4 we show how increasing $\mu$ or going to higher 
order $n$ affects the maximum konus density. As can be seen, at $b/a \gta 1$, 
the central density of the maximum konus density increases with both $\mu$ 
and $n$. For small scalelengths $b$, however, $\Lambda_{\nu}$ is maximal
for the konus density with smallest $\mu$ and $n$ (i.e., $\mu =1$, $n=6$).
In fact the absolute maximum of $\Lambda_{\nu}$ is reached for the konus
density with the smallest allowed values of $\mu$ and $n$.
  
\subsection{4.1 The luminosity profiles}

Any konus density $\rho_{\rm k}(R,z|i_0)$ projects to zero
surface brightness when seen under an inclination angle $i \leq i_0$.
However, when a galaxy to which we add such a konus density is seen
more edge-on ($i > i_0$), it does {\it not} remain invisible.
In order to examine how realistic galaxy models are with a maximum konus
density, we project a perfect oblate spheroid with a maximum
konus density $\rho_{\rm k}(R,z|30\deg)$ under an inclination angle
of $90\deg$ (i.e., we observe this galaxy edge-on), and compare
the luminosity profiles along the major and minor axis with and without
the maximum konus density. The results are shown in Figure 5, where we
plot the logarithm of the projected density versus the logarithm of
the radius from the center projected on the sky $R$. We plot the luminosity
profiles along the major and the minor axis. We consider the $\mu =1$, $n=6$
konus density with $b/a = 1.3$. For this ratio of the characteristic lengths
a maximum value of $\Lambda_{\nu}$ was found (see Figure 3).
%
%
\beginfigure*{7}
\centerline{\psfig{figure=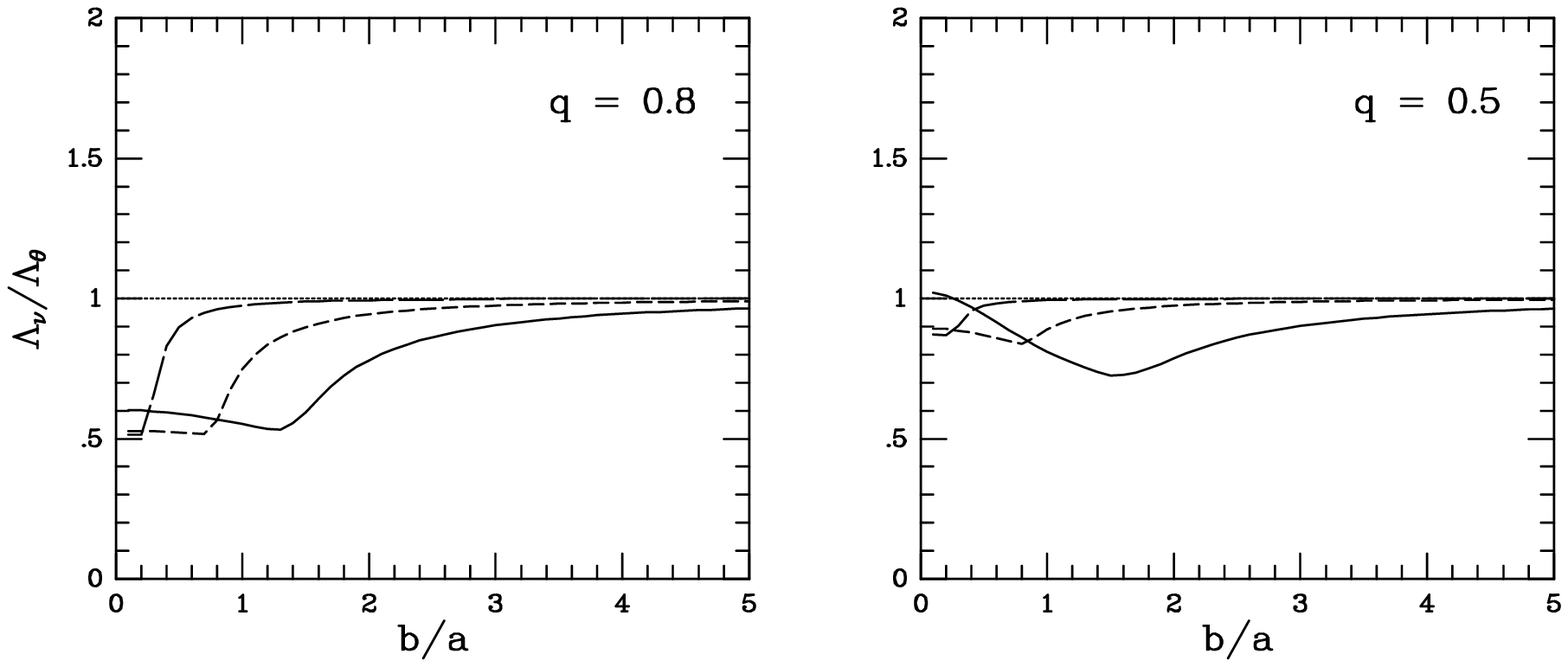,width=\hdsize}}\smallskip
\caption{{\bf Figure 7.} The ratio $\Lambda_{\nu}/\Lambda_{\theta}$ as 
function of the ratio $b/a$ of the $6^{\rm th}$ order generalized konus 
density ($\mu = 1$). The results are presented for three different inclination
angles: $i=30\deg$ (solid lines), $i=60\deg$ (short dashed lines), and 
$i=80\deg$ (long dashed lines). The panel on the left is for a 
perfect oblate spheroid with an ellipticity of 0.2, whereas the 
results in the panel on the right are for a model with 
$\epsilon = 0.5$.} 
\endfigure
%
%

As can be seen the presence of the (maximum) konus density
is clearly visible. Its presence introduces wiggles in the 
luminosity profile along the minor axis. Galaxies with wiggles in their
luminosity profiles are known. However, in all such cases the wiggles
are mainly visible along the {\it major} axis profiles. Examples of
such cases are galaxies that have embedded nuclear disks (see e.g., 
Ferrarese \etal 1994, van den Bosch \& Jaffe 1996). The konus densities 
used here however, reveal themselves by the presence of wiggles along the 
minor axis only. Since no such galaxies are known to exist, we can use 
this as an empirical criterion on the maximum konus density. 
We can make this criterion more qualitative by defining a maximum
amplitude of the wiggle along the minor axis luminosity profile.
We define the maximum konus density according to the empirical criterion,
as that konus density for which 
\eqnam\empcrit
$$\max_{R}\vert\mu_{\rm gal} - \mu_{\rm tot}\vert = 0.05.\eqno\new$$
i.e., for which the maximum difference between the luminosity profiles
with ($\mu_{\rm tot}$) and without ($\mu_{\rm gal}$) konus density is 0.05 
magnitudes. Given the photometric
accuracy with which luminosity profiles can be measured observationally,
any wiggle with an amplitude bigger than 0.05 magnitudes should be
detectable. In Figure 6 we plot the maximum konus density $\Lambda_{\nu}$ 
that can be added to a perfect oblate spheroid with $q=0.8$ as derived by 
using three different criteria: $\rho > 0$ (solid lines), 
$\partial \rho / \partial \nu < 0$ (dashed lines), and the empirical 
criterion (\empcrit) that the minor axis surface brightness profile, when 
projected edge-on, should not have wiggles bigger than 0.05 magnitude 
(dotted lines). The results are shown as a function of $b/a$ for two 
different inclination angles ($i = 30\deg$, and $60\deg$). In almost all 
cases the $\partial \rho / \partial \nu < 0$ criterion is dominant. Only 
for low inclinations, and $b/a \sim 1-3$ is the empirical criterion more 
strict.

%
\beginfigure*{8}
\centerline{\psfig{figure=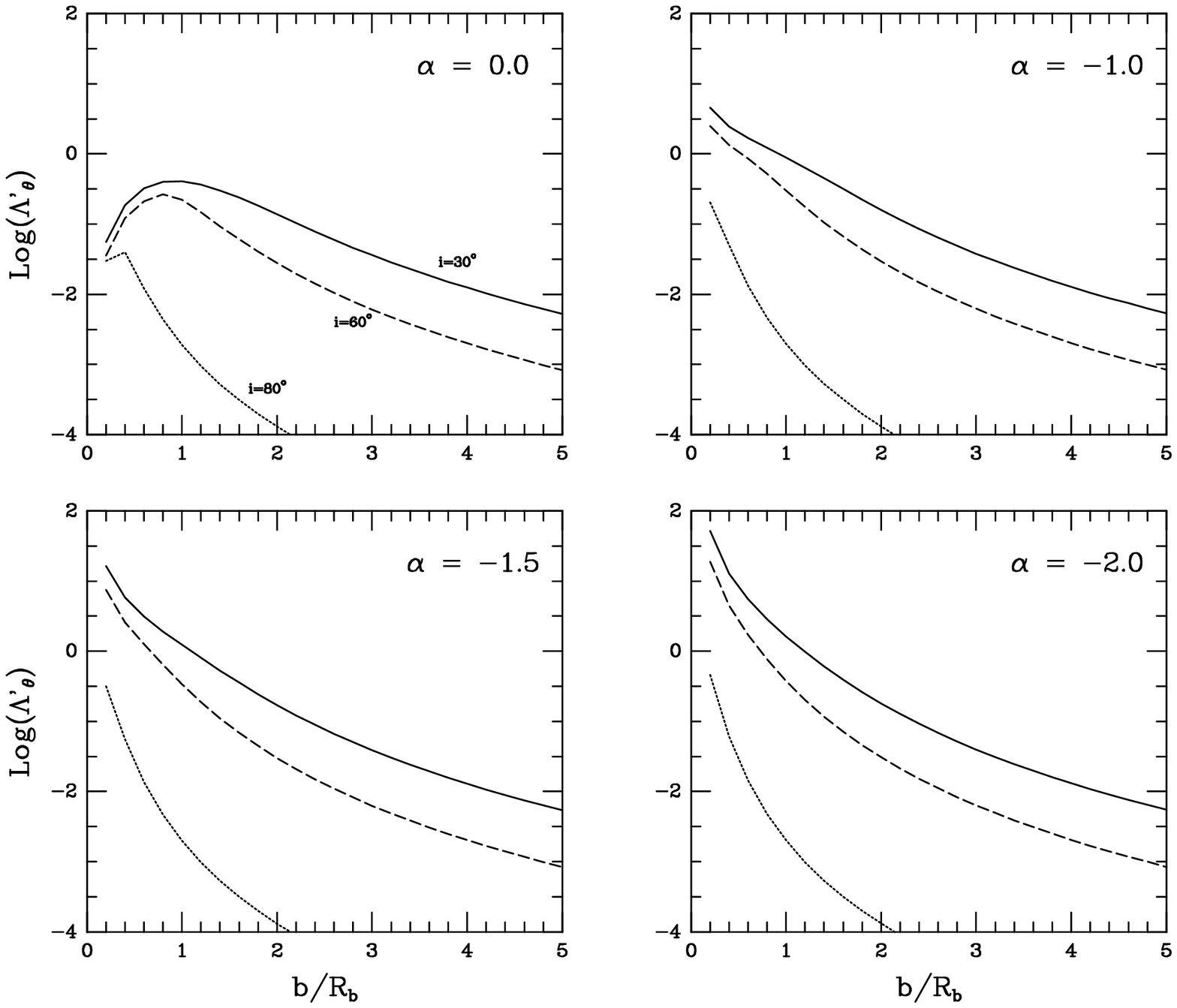,width=\hdsize}}\smallskip
\caption{{\bf Figure 8.} The ratio $\Lambda'_{\theta}$ of the central 
density of the maximum konus density over the density of the 
$(\alpha,\beta)$-model at its core-radius $R_b$ as function of the ratio
$b/R_b$. Results are plotted for four different values of the cusp steepness
$\alpha$, and for three different inclination angles: $i = 30\deg$ 
(solid lines), $i = 60\deg$ (dashed lines), and $i = 80\deg$ (dotted lines).
The maximum konus density increases strongly with increasing cusp steepness
and decreasing konus-scalelength $b$.}
\endfigure
%
%

\eqnumber=1
\def\chaphead{\hbox{5.}}

\section{5 An approximate criterion}

Observations with the Hubble Space Telescope (HST) have revealed that
all elliptical galaxies have central cusps in their density distribution
(e.g., Ferrarese \etal 1994; Lauer \etal 1995). The perfect oblate spheroid
we have studied so far has a density distribution that becomes
constant inside its break-radius $a = \sqrt{-\alpha}$, and therefore is not
a realistic model for an elliptical galaxy. The advantage, however, of the
perfect oblate spheroid is that it is a St\"ackel potential so that we can
use the criterion $\partial\rho/\partial\nu < 0$. Unfortunately, no
cusped St\"ackel models are known. In order to investigate the uncertainty
in the central density of realistic, centrally cusped, elliptical galaxies,
we therefore need to use another criterion. 

We use the criterion that the derivative of the total density
distribution along a circular curve with radius $r = \sqrt{R^2 + z^2}$
has to be negative:
\eqnam\approxcrit
$${\partial\rho \over \partial\theta} < 0,\eqno\new$$
where $\theta = \arctan(z/R)$.
We examine the accuracy of this approximate criterion, by comparing the
values of $\Lambda$ that we find for the perfect oblate spheroid using
both the proper criterion ($\Lambda_{\nu}$) and the approximate criterion 
($\Lambda_{\theta}$). The results are shown in Figure 7, where we plot 
the ratio $\Lambda_{\nu}/\Lambda_{\theta}$ 
as a function of $b/a$ for two different flattenings of the underlying 
perfect oblate spheroid ($q = 0.8$, and $0.5$), and three
different inclination angles ($i=30\deg$, $i=60\deg$, and $i=80\deg$).
The approximate criterion always leads to an overestimate of the
maximum konus density. The error is larger for smaller scalelengths $b$
of the konus density, for smaller inclination angles, and for
less flattened galaxies. 

Having shown that the criterion (\approxcrit) leads to reasonably
accurate approximations of the maximum konus density for the perfect
oblate spheroid, by no means ensures 
that it remains reasonable for, for instance, cusped galaxies.
In fact, the finding that the approximate criterion leads to very accurate
maximum konus densities when the scalelength is large, is somewhat trivial
given that at large radii the prolate spheroidal coordinate
$\nu$ becomes equal to $\theta$. Nevertheless, since the approximate 
criterion enforces a certain smoothness on the density distribution, we feel 
confident that this criterion will lead to more accurate maximum konus 
densities than simply using the criterion that the density has to be 
positive. We showed in Section 2, that under that naive criterion, one 
will overestimate the maximum konus densities (see also Fig. 6). In 
Section 7 we show that the approximate criterion leads to similar maximum 
konus densities as the empirical criterion (\empcrit).
%
\beginfigure{9}
\centerline{\psfig{figure=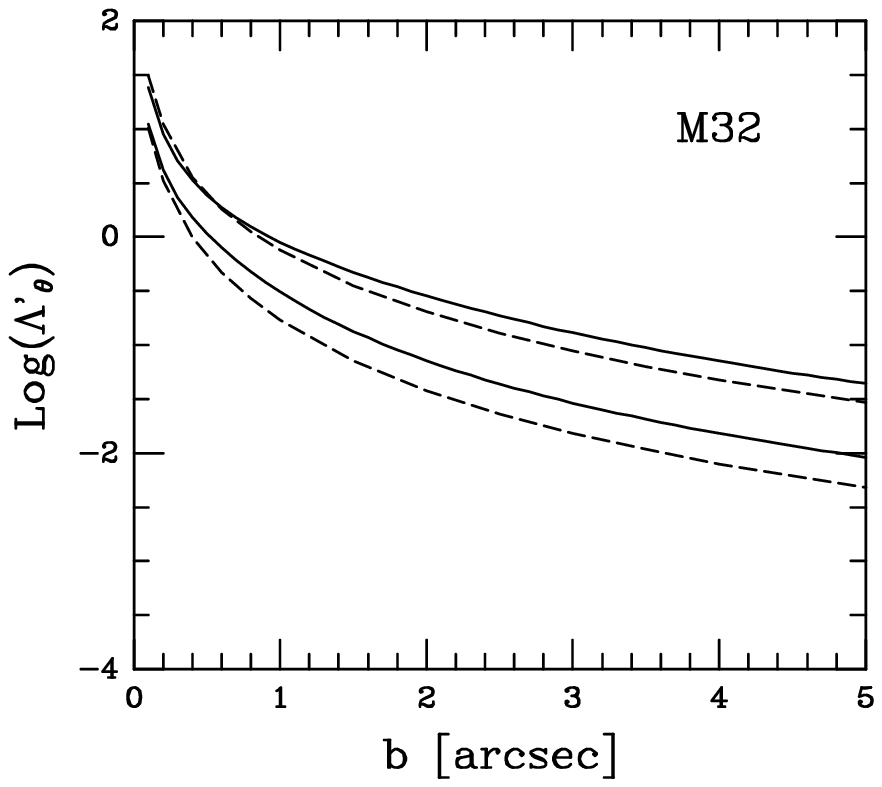,width=\hssize}}\smallskip
\caption{{\bf Figure 9.} The logarithm of the parameter $\Lambda'_{\theta}$ 
as a function of the scalelength $b$ of the generalized konus density for the 
($\alpha,\beta$,$\gamma$)-model of M32. The lower two curves are for the konus
density with $\mu = 1$ and $n=6$,  whereas the upper two curves correspond to
a $\mu =4$, $n=6$ konus density. Solid curves correspond to criterion
(\approxcrit), whereas the two dashed curves show the maximum konus 
density as derived from the empirical criterion (\empcrit).
Both criteria give values for $\Lambda$ of the same order of magnitude, with
the empirical criterion being slightly more strict. }
\endfigure
%
%

\eqnumber=1
\def\chaphead{\hbox{6.}}

\section{6 Cusped density distributions}

We now investigate the maximum konus density one can add to a cusped density 
distribution. We consider the so called
($\alpha,\beta$)-models, which have proven to be useful representations
of elliptical galaxies (e.g., Qian \etal 1995). The density distribution,
which is given by
\eqnam\cuspdens
$$\rho(R,z) = \rho_0 \bigl({m \over R_b}\bigr)^{\alpha} \Bigl(1 +
\bigl({m \over R_b}\bigr)^{2} \Bigr)^{\beta},\eqno\new$$
is stratified on concentric ellipsoids $m=\sqrt{R^2 + z^2/q^2}$ of 
constant flattening $q$. For $R \gg R_b$ they fall off as 
$m^{\alpha + 2\beta}$, while for $R \ll R_b$ they have 
$\rho \propto m^{\alpha}$. We will restrict ourselves to models that 
have $\alpha + 2\beta = -4$ so that at large radii their projected 
surface brightness falls off as $R^{-3}$.

We use the approximate criterion to investigate the dependence of the 
maximum konus density we can add to the density distribution of equation 
(\cuspdens), as function of the cusp steepness $\alpha$. We consider a 
strongly flattened galaxy ($q = 0.5$) seen under three different inclination 
angles ($i = 30\deg$, $i = 60\deg$, and $i = 80\deg$).
When the cusp steepness $\alpha < 0$ the central density of the
($\alpha,\beta$)-model is infinite. We can therefore not calculate the
uncertainty of the central density $\Lambda_{\theta}$. Instead, we
calculate the ratio $\Lambda'_{\theta}$ of the central density of the 
maximum konus density over the density $\rho_0$ of the 
($\alpha,\beta$)-model. We restrict ourselves to a konus density with $\mu=1$
and $n=6$. The results are shown in Figure 8.
When no cusp is present (i.e., $\alpha = 0$) we again find a maximum
$\Lambda'_{\theta}$ at $b/R_b \sim 1$. However, for density distributions that
are cusped we find that $\Lambda'_{\theta}$ keeps increasing
when $b/R_b \rightarrow 0$. In the following
section we will investigate the importance of this for galaxies where
strong indications are found for the presence of a nuclear BH by discussing
a specific example; the compact elliptical M32. 
%
%
\beginfigure*{10}
\centerline{\psfig{figure=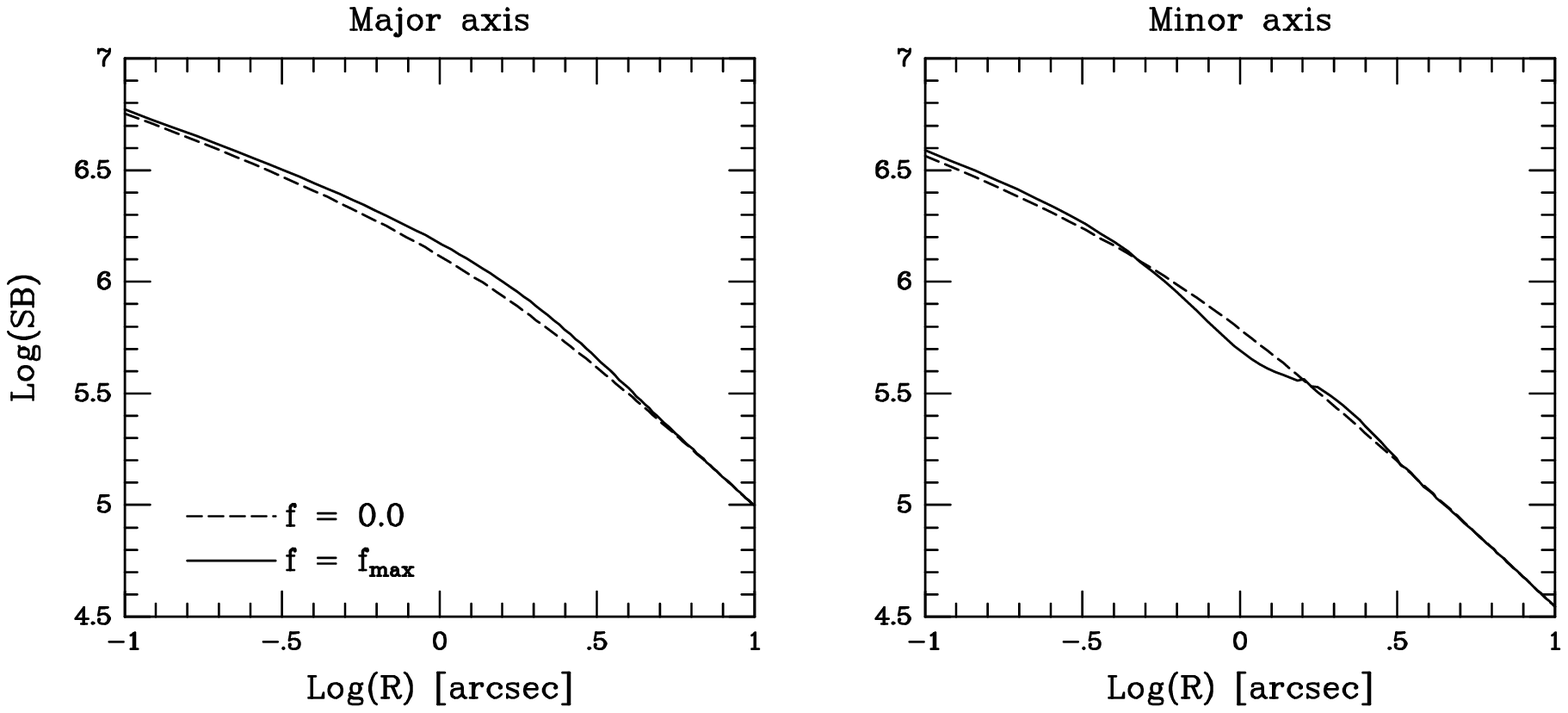,width=\hdsize}}\smallskip
\caption{{\bf Figure 10.} The logarithm of the edge-on projected surface 
brightness (in arbitrary units) of the ($\alpha$,$\beta$,$\gamma$)-model of
M32, both with (solid lines) and without (dashed lines) the maximum konus
density ($\mu =1$, $n=6$) derived from the approximate criterion.}
\endfigure
%
%

\eqnumber=1
\def\chaphead{\hbox{7.}}

\section{7 Application to M32}

To illustrate the importance of konus densities to real galaxies
we here apply our analysis to M32. Strong evidence exists that
M32 harbors a nuclear BH of $\sim 1.8 \times 10^6 \Msun$
(van der Marel \etal 1994, Qian \etal 1995, Dehnen 1995). This evidence is 
based
on the observations of a strong central increase of the velocity dispersion.
In this Section we investigate whether redistributing the central
mass of M32 according to the konus density distribution can 
account for this increase without adopting the presence of a nuclear BH.

Although the inner part
of the galaxy can accurately be described by an ($\alpha$,$\beta$)-model,
this does not fit the outer parts of the galaxy. Multiplication of the
($\alpha$,$\beta$)-model with an additional factor can adequately
describe the strong decrease of the density beyond $\sim 20''$.
We therefore follow the approach taken by van der Marel \etal (1996), and
describe the density distribution of M32 by the following
formula (hereafter referred to as a ($\alpha$,$\beta$,$\gamma$)-model)
\eqnam\cuspdensb
$$\rho(R,z) = \rho_0 \bigl({m \over R_b}\bigr)^{\alpha} 
\Bigl(1 + \bigl({m \over R_b}\bigr)^{2} \Bigr)^{\beta}
\Bigl(1 + \bigl({m \over R_c}\bigr)^{2} \Bigr)^{\gamma}.\eqno\new$$
Projection of this density distribution and fitting to the observed
surface brightness (from both HST and ground-based data) results in a
good fit to the data for the parameters:
$\alpha = -1.435$, $\beta = -0.423$, $\gamma = -1.298$, $R_b=0.55''$,
and $R_c = 102.0''$ (van der Marel, priv. communication). 
Throughout we will assume a distance of M32 of 700 kpc, so that $1''$ 
corresponds to 3.4 pc. 

An important problem with M32 is the fact that the inclination angle
is unknown. Van der Marel \etal (1994) constructed axisymmetric models
of M32 for different assumptions of the inclination angle ($i = 90\deg$ and
$i = 50\deg$).
After solving the Jeans equations they concluded that both models
require the presence of a $\sim 1.8 \times 10^6 \Msun$ BH in order
to fit the observed velocity dispersions in the center. If indeed
M32 is observed edge-on there is no uncertainty in the deprojection
of the surface brightness so that no konus density can be added.
Here we will focus on the $i = 50\deg$ model for which van der Marel \etal
(1994) derived a mass-to-light ratio in the Johnson V-band of 2.36.
Given that the observed flattening of M32 is equal to 0.73, this results
in an intrinsic flattening of the ellipsoids $m$ of 0.452 
(if $i=50\deg$). The density $\rho_0$ of equation (\cuspdensb) is
$1.76 \times 10^5 \Msun {\rm pc}^{-3}$.

We have calculated the maximum konus density $\rho^{n}_{\rm k}$
that we can add to this ($\alpha$,$\beta$,$\gamma$)-model of M32.
Since at large radii $m \propto m^{\alpha + 2\beta +2\gamma} \sim m^{-4.8}$ 
we again need $n \geq 6$. We use the approximate criterion $\partial \rho/
\partial\theta < 0$, and calculate the parameter $\Lambda'_{\theta}$
for different values $\mu$, $n$, and $b$ of the konus density.
The results are shown in Figure 9 for two different values of $\mu$ ($\mu = 1$
and $\mu=4$ both with $n=6$). Unlike with the perfect spheroid, $\Lambda$ is 
larger for higher $\mu$ or $n$ {\it for each value of the scalelength} $b$ 
(compare Figure 4). In principle, we can make $\Lambda'$ arbitrarily large by 
going to small $b$ and large $n$ or $\mu$.  
%
%
\beginfigure{11}
\centerline{\psfig{figure=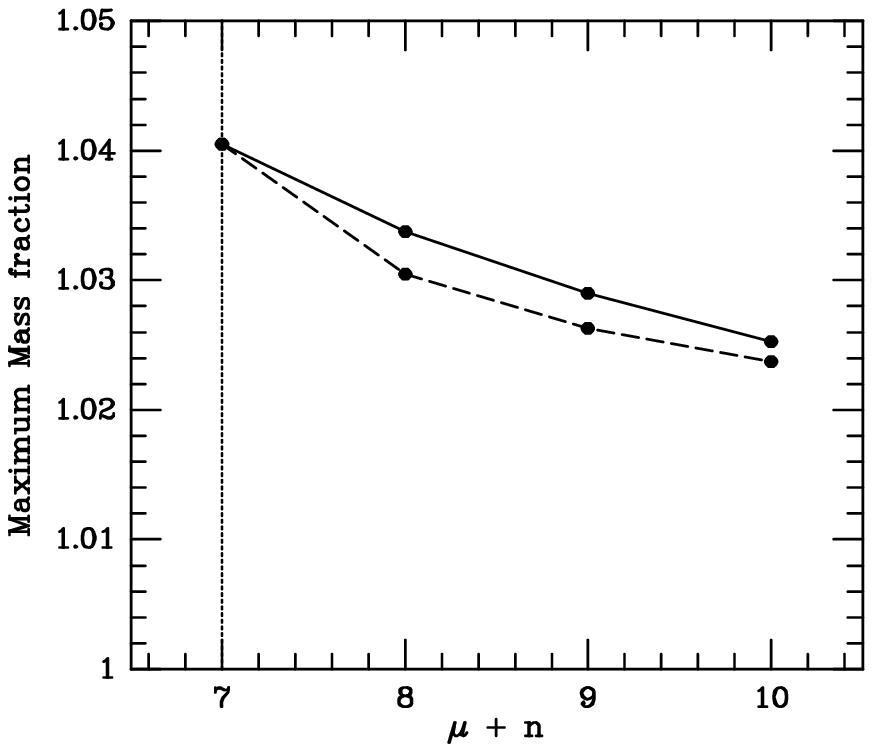,width=\hssize}}\smallskip
\caption{{\bf Figure 11.} The maximum mass fraction $\Gamma_{\rm max}$ 
(equation 7.4) as function of $\mu + n$. Since $\mu \geq 1$ and
$n\geq 6$ we have a minimum $\mu +n = 7$. The solid line shows 
$\Gamma_{\rm kon}$ for increasing order $n$ ($\mu = 1$), whereas the influence
of increasing $\mu$ (with $n=6$) is indicated by the dashed line.}
\endfigure
%
%

In addition, we have determined the
maximum konus density allowed according to the empirical criterion (\empcrit)
that the edge-on projected surface brightness along the minor axis should 
not have wiggles exceeding 0.05 magnitude (dashed lines in Figure 9). 
Both criteria give rather similar values for $\Lambda$, although the 
empirical criterion is more strict. 
This is evident from Figure 10, where we plot the luminosity profiles for the 
($\alpha$,$\beta$,$\gamma$)-model of M32 both with (solid lines) and without 
(dashed lines) the maximum konus as derived from the approximate criterion.

\subsection{7.1 Dynamical significance of konus densities}

We have shown that by increasing $\mu$ or $n$ we can make $\Lambda'$
arbitrarily large. However, $\Lambda'$ is the central konus density divided
by the density of the galaxy at $m=R_b$. Since the central density of a 
cusped galaxy is by definition infinite, it is not {\it a priori} clear
what the real, dynamical significance of these maximum konus densities is.
In order to compare the dynamical significance of the (maximum) konus density
with that of the inferred BH, we have calculated the ratios 
$$\Gamma_{\rm BH}(r) = {M_{\rm gal}(r) + M_{\rm BH} \over M_{\rm gal}(r)},
\eqno\new$$
and
$$\Gamma_{\rm kon}(r) = {M_{\rm gal}(r) + M_{\rm kon}(r) \over M_{\rm gal}(r)}.
\eqno\new$$
Here $M(r)$ is the mass inside a sphere of radius $r$, and $M_{\rm gal}$
is the mass of the ($\alpha$,$\beta$,$\gamma$)-model. Although the total
mass of the konus density is zero, at sufficiently small radii
there is only a positive density contribution resulting in a non-zero
positive $M_{\rm kon}(r)$. Since the central konus density is finite,
$\Gamma_{\rm kon}(r)$ is zero at $r=0$, reaches a maximum at 
a certain intermediate radius, and decreases to zero again
at larger radii. This $\max_{r} \Gamma_{\rm kon}(r)$ depends on the scalelength
$b$ of the konus density, as well as on $\mu$ and $n$. For each $\mu$ and $n$
we calculate the maximum mass fraction, defined as
\eqnam\massfrac
$$\Gamma_{\rm max} = \max_{b}\Bigl[\max_{r} \Gamma_{\rm kon}(r)\Bigr],
\eqno\new$$ 
This maximum mass fraction as function of $\mu + n$ is shown in Figure 11. 
Although increasing $\mu + n$ increases the central density of the maximum
konus density (see Figure 9), the number of oscillations increases as well.
Therefore, the central region where $\rho_{\rm k}>0$ becomes smaller and we
find that $\Gamma_{\rm max}$ {\it decreases} with increasing $\mu+n$.
For our generalized konus density we find an absolute maximum mass fraction
of $\sim 1.04$ only. This maximum is reached for a konus density with 
$b=0.22''$, $\mu = 1$, and $n=6$. It is not only important what the absolute
mass fraction is, but also at what radius it is reached.  In Figure 12 we 
have plotted the radial run of $\Gamma_{\rm BH}$. If $\Gamma_{\rm max} = 1.04$
is reached at 5 pc, it is comparable to $\Gamma_{\rm BH}$. However, the
maximum mass fraction occurs at very small radii ($r \approx 0.7$pc), where
$\Gamma_{\rm BH}$ is many orders of magnitude larger. If the scalelength $b$
of the konus density increases, the radius where $\Gamma_{\rm kon}(r)$ is
maximal increases, but since $\Lambda$ decreases strongly with scalelength
(Figure 9), the maximum $\Gamma_{\rm kon}$ also decreases with $b$.
We therefore conclude that the dynamical influence of the maximum 
generalized konus density is negligible compared to that of the inferred BH.
%
%
\beginfigure{12}
\centerline{\psfig{figure=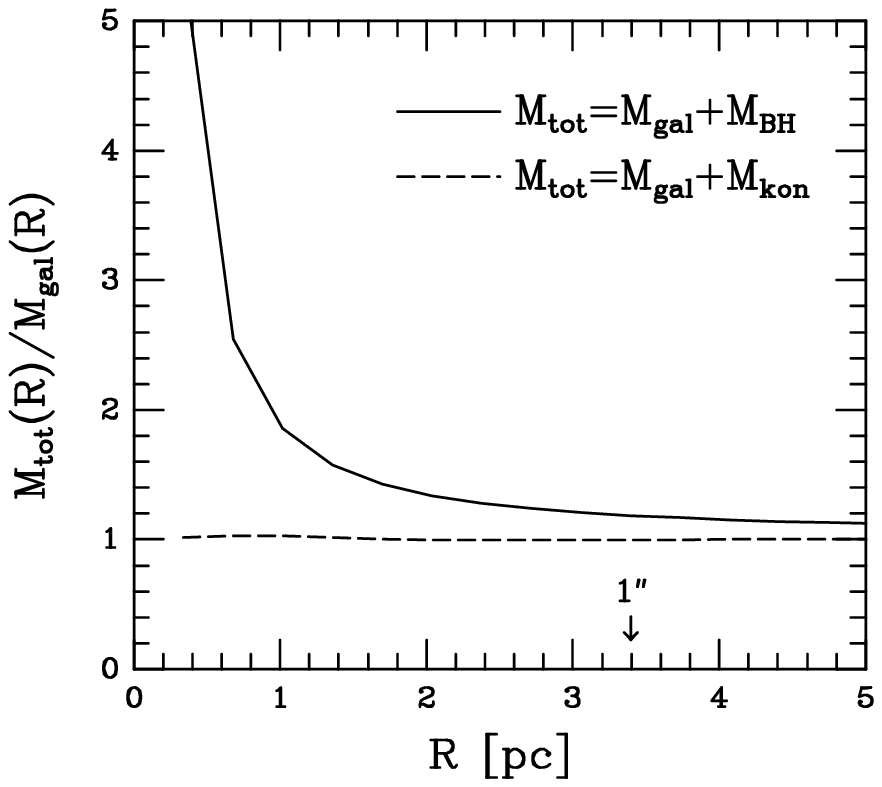,width=\hssize}}\smallskip
\caption{{\bf Figure 12.} The total mass inside a sphere of radius $R$, 
$M_{\rm tot}(R)$, over the mass of the M32 density model
inside the same sphere as function of radius $R$. The solid line shows
the case in which the total mass is build up of the 
($\alpha,\beta$,$\gamma$)-model of M32 with in addition a BH of 
$1.8 \times 10^6 \Msun$. The dashed line is for the case in which we have 
added the maximum generalized konus density with $b=1.0''$ to the 
density model of M32.}
\endfigure
%

\subsection{7.2 Cusped konus densities}

The Fourier Transform of the generalized konus density decays exponentially 
at large frequency $k$ (see Appendix B). This enforces $\partial \rho_k / 
\partial r$ to go to zero at small radii $r$. Here we investigate whether
konus densities with a central density cusp can add more mass to the center.
In order for a konus density to be cusped, i.e., $\rho_{\rm k} \propto 
r^{-\alpha}$ ($0 < \alpha < 3$) at small radii, its Fourier Transform should 
decay as $k^{\alpha - 3}$ at large frequencies $k$.
The requirement $\alpha < 3$ ensures that the density cusp has finite mass. 
We can constrain this regime even further by taking into account the fact
that every konus density projects to zero surface brightness when
projected face-on (i.e., with $i=0\deg$). Since
$$\int_0^{a} r^{-\alpha} {\rm d}r = \infty,\eqno\new$$
for $\alpha \geq 1$ and $a>0$, the projected surface brightness at
the center can only be equal to zero if $\alpha < 1$. Therefore, konus
densities can at most be moderately cusped, with a central density
gradient less steep that $r^{-1}$.
%
%
\beginfigure{13}
\centerline{\psfig{figure=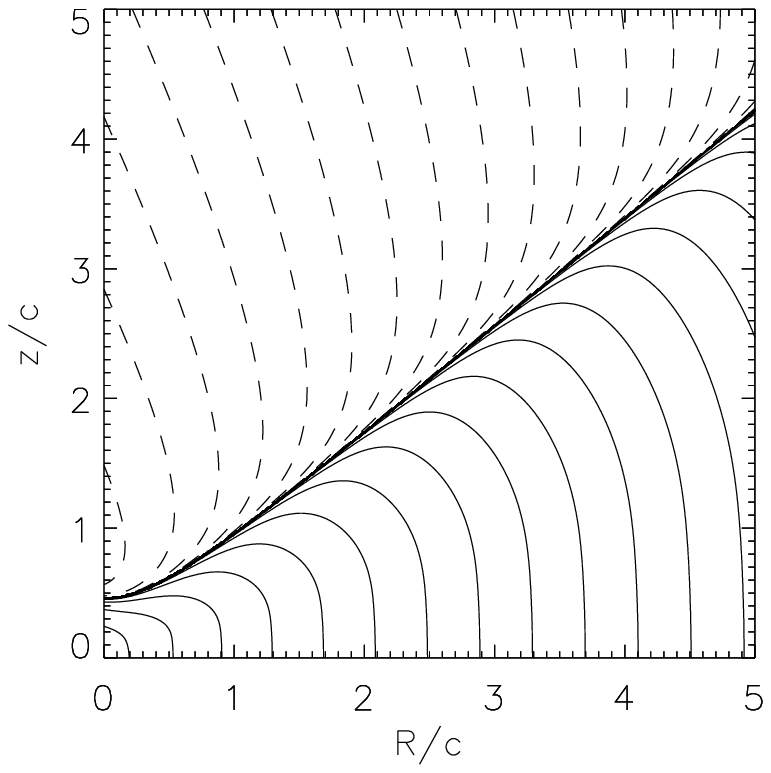,width=\hssize}}\smallskip
\caption{{\bf Figure 13.} A contour plot of the cusped `konus'-density
$\rho_c(R,z|50\deg)$ for $\alpha = 0.25$. The parameters $b$ and $d$ are
taken such that $\rho_c(R,z|50\deg)$ obeys the criteria (7.7) and
(7.8).}
\endfigure
%
 
As we have seen in the previous section, maximizing $\Gamma_{\rm kon}$, not 
only requires high central density, but also a sufficiently large
area in the center where $\rho_{\rm k} > 0$.
A cusped konus density with this property, in order to have zero total mass,
must have regions outside the center with strongly negative density.
Addition of such a konus density, is therefore likely to result in
luminosity profiles (when projected edge-on) with large wiggles. In other
words, it seems likely that the empirical criterion (\empcrit) becomes
most strict. In fact, already for our generalized konus density, this criterion
is the most restricting (see Figure 9). In order to quantify this, we have 
tried to construct cusped konus densities, but were unable to find an 
analytical $\rho_{\rm k}(R,z)$ that obeys the physical
restrictions of konus densities, i.e., zero total mass, sufficiently rapid 
decay at large radii, and no discontinuities. We therefore constructed a 
density distribution that {\it resembles} a cusped konus density. 
Consider the density distribution
\eqnam\fakekon
$$\rho_{c}(R,z|i) = $$
$$\; \; \; \rho_0 \; r^{-\alpha} \; e^{-r/c} \; e^{-R/b} 
\tanh \Bigl({\sqrt{R^2 + d^2}\tan^{-1}i - z\over z}\Bigr).\eqno\new$$
Here $r=\sqrt{R^2 + z^2}$, and $\alpha$, $b$, $c$, and $d$ are free parameters.
This density is cusped in the center (with power law slope $\alpha$),
and decays exponentially at large radii $r >> c$. At 
$z=\sqrt{R^2 + d^2}\tan^{-1}i$ the density changes sign, becoming negative at
larger $z$. We adopt $\alpha$ and $c$ as free parameters and use the
physical properties of genuine konus densities to determine $b$ and $d$.
Since we have added the factor $e^{-R/b}$, which equals unity
for $z=0$, we can use the requirement that the line-of-sight integral
along the minor axis should yield zero, i.e.,
\eqnam\loseqzer
$$\int_{-\infty}^{\infty} \rho_{c}(R=0,z|i) dz = 0,\eqno\new$$
to numerically evaluate the ratio $d/c$. Once $d$ is known, we determine
the ratio $b/c$, by numerical evaluation of the root of 
\eqnam\totmaszer
$$M_{\rm tot} = 4 \pi \int_0^{\infty} {\rm d}R \; R
\int_0^{\infty} {\rm d}z \; \rho_{c}(R,z|i) = 0.\eqno\new$$
A contour plot of a cusped `konus'-density with $\alpha=0.25$ that obeys 
these criteria is shown in Figure 13.
Although (\fakekon) is not a konus density, we can interpret it as such and 
use the empirical criterion (\empcrit) to determine the maximum 
`konus'-density. The empirical criterion uses the edge-on projected 
surface brightness along the minor axis, and along this
axis (\fakekon) at least resembles a genuine cusped konus density, in that
it obeys criterion (\loseqzer). As we have seen for the generalized konus
density, the maximum mass fraction decreases for an increasing number
of density oscillations. Our cusped `konus'-density has a minimal number
of density oscillations, and is therefore biased towards larger 
$\Gamma_{\rm kon}$. We have calculated, as function of $\alpha$ and scalelength
$c$, the $\max_{r} \Gamma_{\rm kon}(r)$ for our cusped `konus'-density when
added to M32. The results are shown in Figure 14. Similar as for the 
generalized konus density, $\max_{r} \Gamma_{\rm kon}(r)$ increases with
decreasing scalelength. More importantly, increasing 
the cusp slope $\alpha$ {\it decreases} $\Gamma_{\rm kon}$. This is due
to the fact that for larger $\alpha$ the konus-criterion (\loseqzer)
ensures that $d$ decreases, so that the central region where $\rho_c > 0$
becomes smaller. Evidently, this effect is stronger than the increase in
cusp slope. We therefore conclude, that even cusped konus densities
have a negligible dynamical effect on the central region of M32, as compared
to that of the inferred BH mass. In the analysis above we have used a BH mass
of $1.8 \times 10^6 \Msun$. More recent studies hint towards a more massive
BH of $\sim 3 \times 10^6 \Msun$ (Bender, Kormendy \& Dehnen 1996; 
van der Marel \etal 1996). This only strengthens the conclusions outlined above.
%
%
\beginfigure{14}
\centerline{\psfig{figure=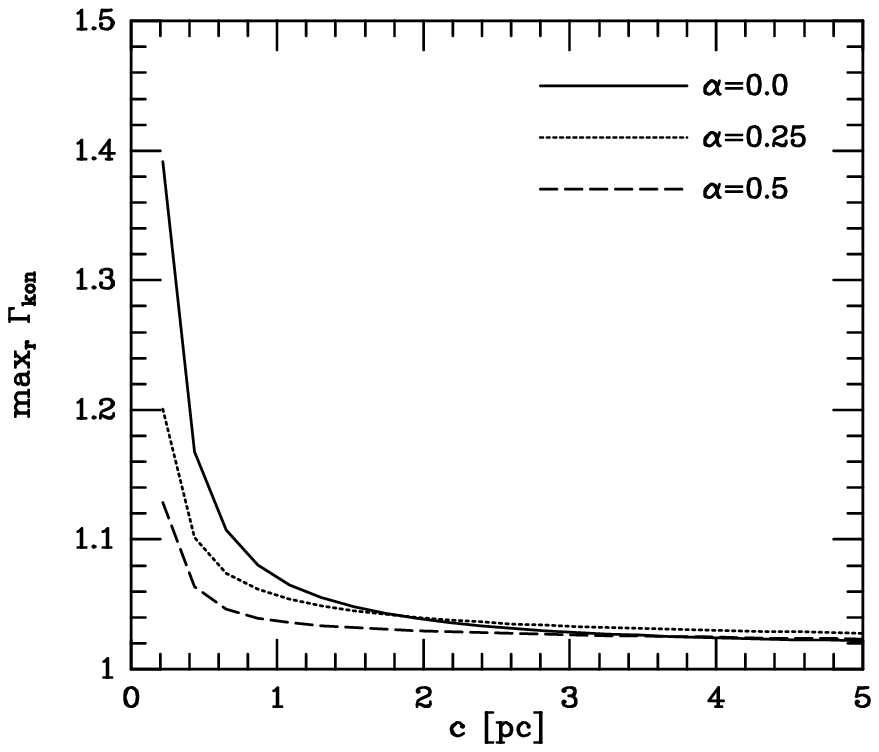,width=\hssize}}\smallskip
\caption{{\bf Figure 14.} The maximum of $\Gamma_{\rm kon}$ over radius $r$
for our cusped `konus'-density $\rho_c$ (\fakekon) as function of
its scalelength $c$, and cusp slope $\alpha$. Increasing the cusp slope,
decreases the mass fraction added to the center by the `konus'-density,
therewith resulting in a smaller dynamical effect as compared to that
of the BH.}
\endfigure
%

\section{8 Conclusions}

The deprojection of axisymmetric systems is indeterminate when
the inclination angle $i \neq 90\deg$. To any such system one
can add densities which, after projection, result in zero surface brightness.
Since these density distributions have zero total mass, addition or
subtraction of konus densities is similar to redistributing the
mass of the system. In this paper we have investigated to what extent
one can redistribute the mass in elliptical galaxies. In particular we
investigated the uncertainties of the central densities of such galaxies
due to the non-uniqueness of the deprojection.

Since konus densities have regions with both positive and negative density,
there is a limit on how much konus density can be added to any
density distribution while maintaining positivity. Although this criterion
will result in a maximum konus density, it ignores whether a model
which such a maximum konus density can actually be built from its orbit
building blocks. We therefore use a more realistic (and more strict)
criterion to determine the maximum konus density.

We used both a physical and an empirical criterion. The physical criterion
is based on the principle that the density should be monotonically decreasing
along curves that follow thin tube orbits. This criterion, although 
sufficient is not strict. For St\"ackel potentials, these thin tube orbits
are described by the curves of constant prolate spheroidal coordinate 
$\lambda$, and the criterion can be made quantitative. For any other potential,
no analytic description of the curves is known. In those cases we use
an approximate criterion that states that
$\partial\rho/\partial\theta < 0$ along circles of constant radius 
$r = \sqrt{R^2 + z^2}$ with $\theta = \arctan(z/R)$.
The empirical criterion we used is based on the fact that adding too much
konus density will result in wiggles along the minor axis luminosity profile,
when the galaxy is seen edge-on. Since no such systems have ever been 
observed this puts an empirical constraint on the maximum konus densities.
  
We calculated the ratio of the central density of the maximum konus density
to the central density of the perfect oblate spheroid to which that konus 
density can be added. For sufficiently small inclination angles 
one can add konus densities whose central density is comparable to that of the
perfect oblate spheroid. For highly inclined galaxies the uncertainty of the 
central density is small however: $\lta 10$\% for $i \gta 80\deg$. 
Perfect oblate spheroids have a break radius inside which the density is
constant. Elliptical galaxies, however, are known to harbor central density
cusps. We therefore applied the approximate criterion to cusped, 
axisymmetric models that have $\rho \propto r^{-\alpha}$ at small radii. 
The steeper the cusp, the more konus density one can add. The central density 
of the maximum konus density increases strongly with decreasing scalelength 
of the konus density and with decreasing inclination angle.

In order to better understand the dynamical significance of maximum konus
densities, we applied our analysis to the specific example of M32; a compact
elliptical for which strong indications are found for the presence of
a central BH of $\sim 2 \times 10^6 \Msun$. The central density distribution
of M32 is strongly cusped ($\rho \propto r^{-1.435}$), and we used both the 
approximate criterion and the empirical criterion to determine the maximum 
amount of konus density. The latter was found to be more strict.
Although for sufficiently small scalelengths of the konus density its
central density can be very large, its dynamical influence on the central
region of M32 is negligible. This is due to the cusped nature of the central 
density distribution of M32. 

We have also investigated the dynamical influence of {\it cusped} konus 
densities.
We have shown, based on a simple argument, that konus densities can only
be moderately cusped ($\rho_{\rm kon}(r) \propto r^{-\alpha}$ with 
$\alpha < 1$). We were unable to find an analytic, cusped konus density,
and therefore used a cusped density distribution that {\it resembles} a
cusped konus density. We made sure that this density distribution has zero
total mass, and when projected face-on, has zero surface brightness in the
center. Using the empirical criterion, we showed that for increasing 
cusp slope $\alpha$, the mass fraction that the konus adds to the center
{\it decreases}. This is due to the fact that an increase in cusp slope,
results in a decrease of the central region where $\rho_{\rm kon} > 0$.

We conclude therefore that konus densities play only a marginally
important role in the study of the dynamics of the central regions of
(cusped) elliptical galaxies. Although considerable amounts of konus density
can be added, when its scalelength is sufficiently small, its dynamical
influence on the central dynamics remains negligible. Especially, we conclude
that the uncertainties in the central mass of axisymmetric galaxies due to
the non-uniqueness of the deprojection are sufficiently small that they
cannot exclude the requirement of a central BH based on the finding of too 
small central densities in these galaxies. This is more and more true 
in galaxies seen under larger inclination angles.

A major shortcoming in the investigations discussed in this paper is the
fact that only one particular class of konus densities was used.
However, the different free parameters of this generalized konus density
allowed a wide range of konus densities to be investigated. Although the
cusped `konus' density distribution was not a real konus density, it at
least obeyed two konus criteria, and we merely used it to investigate
the {\it changes} in mass fraction when cusp slope is increased, rather than
to interpret the {\it absolute} mass fraction for this `konus' density.
We are therefore confident that the conclusions reached in this paper are at
most very moderately dependent on our actual choice of konus densities used.

\section*{Acknowledgments}

\tx The author would like to thank Nicolas Cretton, Walter Jaffe, Tim de
Zeeuw, and Roeland van der Marel for valuable discussion. In addition, the
author is grateful to the referee, Dr. P.L. Palmer, whose comments helped to 
improve the paper considerably. This research 
was financially supported by the
Netherlands Foundation for Astronomical Research (ASTRON), \#782-373-055,
with financial aid from the Netherlands Organization for Scientific Research 
(NWO).

\section*{APPENDIX A: The perfect oblate spheroid}

Kuzmin (1953; 1956) showed that their exists exactly one oblate potential for
which the density is stratified on similar, concentric aligned spheroids,
and for which the equations of motion are separable. This model has become
known as the perfect oblate spheroid (de Zeeuw \& Lynden-Bell 1985), and has 
a potential of St\"ackel form, i.e., 
$$\Phi = - {(\lambda + \gamma) G(\lambda) - (\nu + \gamma) G(\nu) \over
\lambda - \nu} ,\eqno({\rm A1})$$
where $G(\tau)$ is an arbitrary function $(\tau = \lambda , \nu)$,
and $(\lambda,\phi,\nu)$ are prolate spheroidal coordinates, in which the
equations of motion are separable.
The coordinates $\lambda$ and $\nu$ are defined as the roots
for $\tau$ of
\eqnam\prolsphcoor
$$ {R^2 \over \tau + \alpha} + {z^2 \over \tau + \gamma} = 0,\eqno({\rm A2})$$
where $R$ and $z$ are the cylindrical coordinates, $\alpha$ and $\gamma$ are 
negative constants, and we choose
$-\gamma \leq \nu \leq -\alpha \leq \lambda$. The foci of the coordinates
are at $(R,z) = (0,\pm\Delta)$, with $\Delta^2 = \gamma - \alpha$.
The coordinates $(R,z)$ can be written as
\eqnam\cylcoor
$$R^2 = {(\lambda + \alpha)(\nu + \alpha) \over \alpha - \gamma},\;\;\;
  z^2 = {(\lambda + \gamma)(\nu + \gamma) \over \gamma - \alpha}.
\eqno({\rm A3})$$

The density distribution of the perfect oblate spheroid is given by
\eqnam\pscyldens
$$ \rho(R,z) = \rho_0 {a^4 \over (a^2 + m^2)^2},\eqno({\rm A4})$$
where $a = \sqrt{-\alpha}$, and $m^2 = R^2 + z^2/q^2$ are the 
spheroids with flattening $q=\sqrt{\gamma/\alpha}$. The density distribution in
the equatorial plane has a break radius at $a$.
At radii $r >> a$ the density falls of as $r^{-4}$ (consistent with the 
fact that the surface brightness of most galaxies falls off as 
approximately $R^{-3}$). At small radii ($r << a$) the density is constant
and equal to the central density $\rho_0$. The density
distribution can be expressed in prolate spheroidal coordinates as
\eqnam\pspscdens
$$ \rho(\lambda,\nu) = \rho_0 \biggl({\alpha\gamma \over \lambda\nu}\biggr)^2.
\eqno({\rm A5})$$
%
%
\beginfigure*{15}
\centerline{\psfig{figure=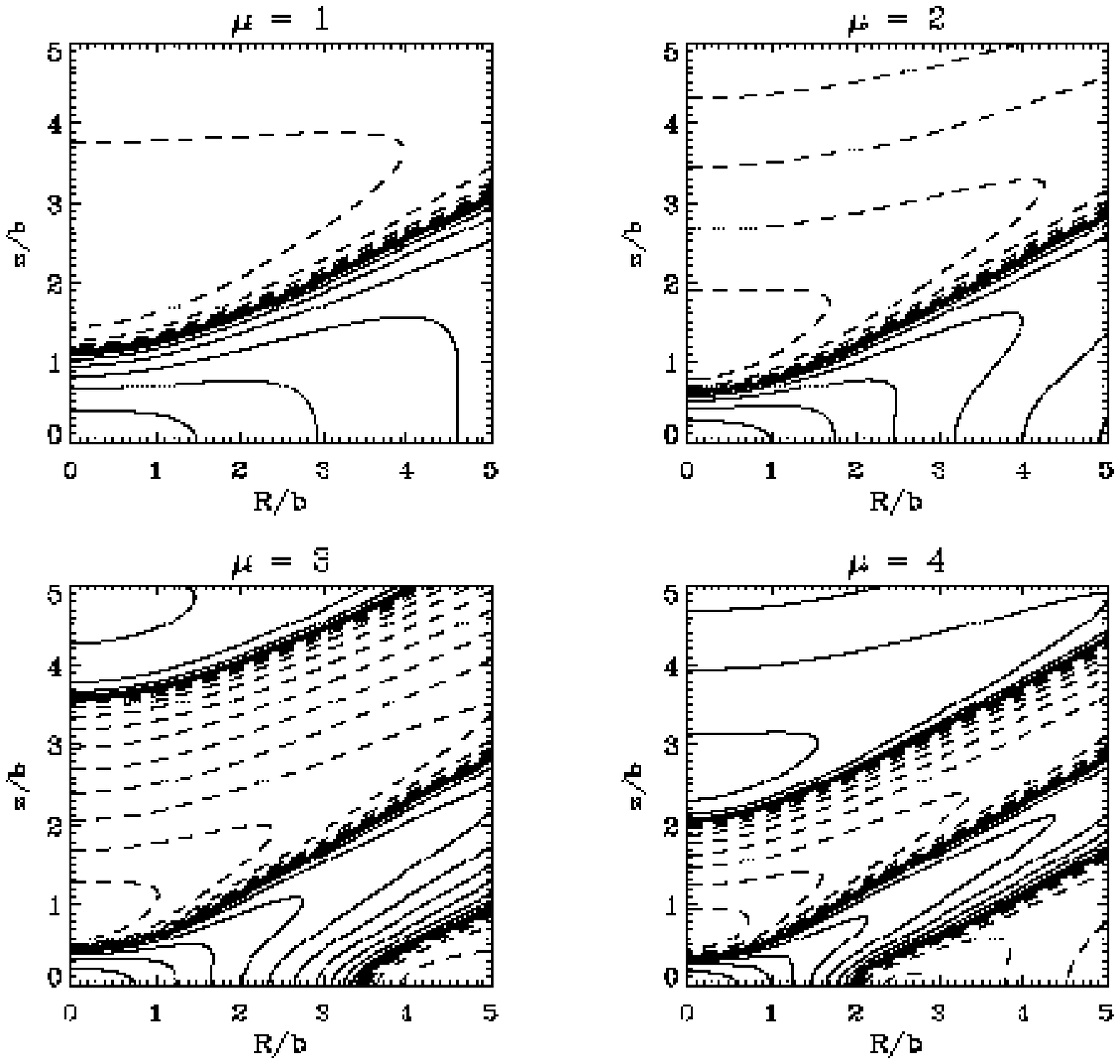,width=\hdsize}}\smallskip
\caption{{\bf Figure 15.} Contours of the generalized $\rho_{k}(R,z|60\deg)$
for $\mu=1,2,3,$ and $4$. Each of these density distributions projects
to zero surface brightness for $i<60\deg$. Positive contours are solid,
negative contours are dashed.}
\endfigure
%
%

\section*{APPENDIX B: A generalized konus density}

Here we follow the approach taken by KR, and consider the class of semi-konus
densities which can be written as
\eqnam\semikon
$$ \rho_{\rm sk}(R,z|i) = 
{[f_s(\alpha_{+}) + f_s(\alpha_{-})] \over r} - 
i\; {[f_c(\alpha_{+}) - f_c(\alpha_{-})] \over r},\eqno({\rm B1})$$
where $\alpha_{\pm} = \sqrt{R^2 + z^2} \cos i \pm z$, and $f_s(x)$ and
$f_c(x)$ are odd and even functions, respectively, defined by
\eqnam\fcx
$$f_c(x) = {1\over 4\pi^2} \int\limits_{0}^{\infty} {\rm d}k \; g(k) \;
\cos(kx),\eqno({\rm B2})$$
and
\eqnam\fsx
$$f_s(x) = {1\over 4\pi^2} \int\limits_{0}^{\infty} {\rm d}k \; g(k) \;
\sin(kx).\eqno({\rm B3})$$
Upon choosing a weight function $g(k) \propto k^{\mu -1} e^{-bk}$ one derives
with the help of formulae 3.944.5 and 3.944.6 in Gradshteyn \& Ryzhik (1980):
\eqnam\fcxex
$$f_c(x) =  {\Gamma(\mu) \over (b^2 + x^2)^{\mu/2}}
\cos\Bigl[\mu \; \arctan\big({x\over b}\big)\Bigr],\eqno({\rm B4})$$
and
\eqnam\fsxex
$$f_s(x) = {\Gamma(\mu) \over (b^2 + x^2)^{\mu/2}}
\sin\Bigl[\mu \; \arctan\big({x\over b}\big)\Bigr],\eqno({\rm B5})$$
where $\Gamma(x)$ is the gamma function.
These functions define a continuous class of konus densities
$\rho_{\rm k}(R,z|i) = \Re \rho_{\rm sk}(R,z|i)$ for $\mu > 0$.
Here $\Re f$ denotes the real part of $f$. For $\mu=1$ the semi-konus 
density (B1) reduces to the first order semi-konus density given by
KR in their equation (20). In the following we will only 
consider integer values of $\mu$. 

In the limit $r = \sqrt{R^2 + z^2} \rightarrow \infty$ the
konus densities $\rho_{\rm k}(R,z|i)$ decline as
$r^{-(\mu+1)}$ for $\mu$ odd and as $r^{-(\mu+2)}$ for $\mu$ even.
This continuous set of konus densities therefore can be chosen to
decline arbitrary rapidly at large radii. In Figure 15 we show contour plots
for the cases with $\mu = 1,2,3,$ and $4$. Note that for $\mu \geq 3$
these konus densities also change sign in the equatorial plane.

As we prove in Appendix C, the generalized konus density 
$\rho_{\rm k}^{\mu}(r,z|i_0)$ with $\mu = 1$ has infinite total mass. 
For $i < i_0$ it projects to constant, but {\it non-zero} surface brightness.
Therefore, addition of such a konus density is no longer equivalent to simply 
redistributing the density. For $\mu = 2$ the
total mass is finite but positive, whereas for $\mu > 2$ the total mass is
zero (see Appendix C). 

As was shown by KR, any function
\eqnam\highordkon
$$\rho^{n}(R,z) = \Re \bigg[\rho_{\rm sk}^{1}(R,z)\bigg]^{n},\eqno({\rm B7})$$
is a konus density as long as $\rho_{\rm sk}^{1}(R,z)$ is a semi-konus
density. KR have presented examples of such higher order konus densities
for the case of $\mu = 1$ of our generalized konus density.

As discussed in the section 3, in order to ensure that realistic
axisymmetric galaxy models can be constructed with the addition or subtraction
of konus densities, one needs konus densities that decline sufficiently fast.
Any konus density of the form (B1), with $f_{s}(x)$ odd 
($f'_{s}(0) \ne 0$) and $f_{c}(x)$ even ($f_{c}(0) \ne 0$) has 
$\rho^n_k \propto r^{-n}$ for $n$ even, and $\partial \rho^n_{\rm k} /
\partial \theta \propto r^{-(n-1)}$ for $n$ odd along lines $\alpha_{\pm} = 0$.
Here $\theta = \arctan(z/R)$, but the same is true for the derivative with 
respect to the prolate spheroidal coordinate $\nu$. In order to have a konus 
density that declines faster than $r^{-4}$ at large radii, and whose derivative 
$\partial \rho_{\rm k}/\partial\theta$ declines faster than $r^{-4}$, we need
to go to order $n\geq 6$. Besides a more rapid asymptotic decline for
higher order $n$, the konus density will also show a greater number of
sign alterations. 
%
%
\beginfigure{16}
\centerline{\psfig{figure=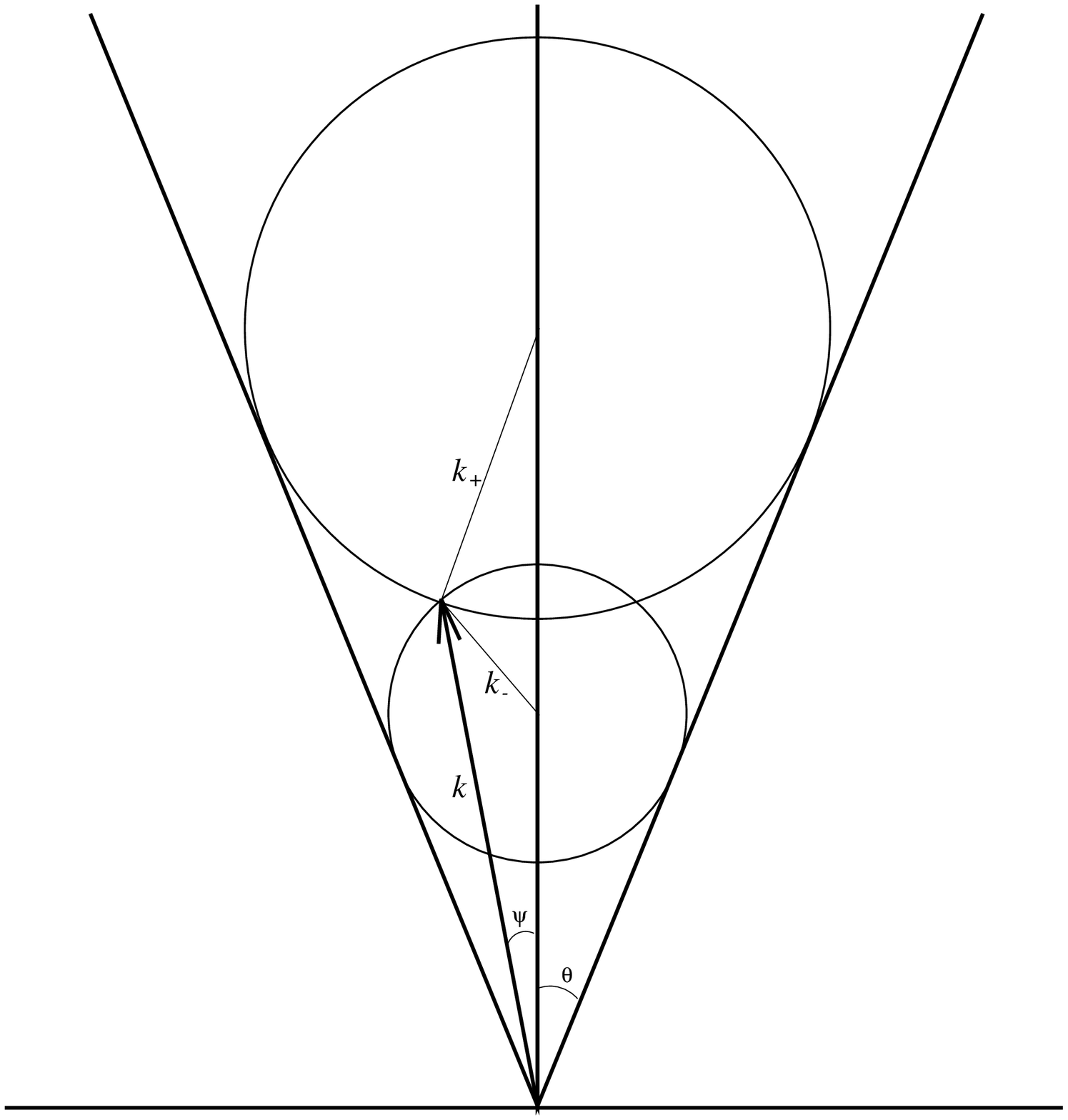,width=\hssize}}\smallskip
\caption{{\bf Figure 16.} The build-up of the Fourier
Transform $H(k,\psi)$ of the semi-konus density inside a cone with
half-angle $\theta$, by a weighted sum of spheres. At each point $(k,\psi)$
with $\psi < \theta$ there are two spheres with radii $k_{+}$ and $k_{-}$ 
that contribute to $H(k,\psi)$.}
\endfigure
%
%

\section*{APPENDIX C: The total mass of the generalized konus densities}

Let $H(\vec k)$ be the Fourier Transform of the semi-konus density
$\rho_{\rm sk}(r,z|\theta)$, where $\theta = 90\deg - i$. We will use
spherical coordinates $(k,\psi,\phi)$ for the Fourier space vector $\vec k$.
Then $H(\vec k) = H(k,\psi)$ since the Fourier Transform of the semi-konus 
density is symmetric around the $k_z$ axis.
We follow the approach taken by KR and build up $H(\vec k)$ by
a weighted sum of kernel functions $f(\vec k, \vec k_0)$, i.e.,
$$H(\vec k) = \int\limits_0^{\infty} {\rm d}k_0 \; g(k_0) 
f(\vec k, \vec k_0).\eqno({\rm C1}) $$
For our specific semi-konus density of the form (B1), the kernel is a 
shell of radius $\left|\vec k_0\right| = k_0$ shifted along the $k_z$ axis 
by an amount $k_0/\sin\theta$, and 
$$f(\vec k, \vec k_0) = {\delta\bigl[\left|\vec k - \vec k_0/\sin\theta\right|
- \left|\vec k_0\right|\bigr] \over \left|\vec k_0\right|},\eqno({\rm C2}) $$
where $\delta(x)$ is the delta function. This is illustrated in 
Figure 16, where we have plotted the cone with half-angle $\theta$ in the 
meridional plane of $\vec k$.
As can be seen, at any point $(k,\psi,\phi)$ (with $\psi < \theta$)
there are two shells that contribute weight to that point. 
We define the function $f(k_0)$ as
$$\eqalign{f(k_0) 
&\equiv \left|\vec k - \vec k_0/\sin\theta\right| - \left|\vec k_0\right|\cr
&= \sqrt{k^2 + {k_0^2\over\sin^2\theta} - 2 k {k_0\over\sin\theta} \cos\psi} 
- k_0,\cr}\eqno({\rm C3}) $$
From the normalization relation of $\delta$-functions we have that
$$\delta\bigl[f(k_0)\bigr] 
= {\delta(k_0 - k') \over f'(k')},\eqno({\rm C4})$$
where $k'$ are those $k_0$ for which $f(k_0) = 0$.
Solving $f(k_0) = 0$ we find two solutions $k' = k_{\pm}$
$$k_{\pm} = k {\sin \theta \over \cos^2 \theta} \biggl(\cos \psi \pm
\sqrt{\cos^2\psi - \cos^2\theta}\biggr).\eqno({\rm C5}) $$
Calculating the derivative $f'(k_0)$ and using equations (C2), (C4), and (C5)
we find (for $\psi < \theta)$
$$H(\vec k) = {\sin\theta \over \sqrt{\cos^2\psi - \cos^2\theta}}
\Biggl[ {g(k_{+})\over k} - {g(k_{-})\over k}\Biggr].\eqno({\rm C6}) $$

The function $H(\vec k)$ can also be written as the Fourier Transform of
the semi-konus density, i.e.,
$$H(\vec k) = \int {\rm d}^3\vec r \; \rho_{\rm sk}(\vec r|\theta) \;
\exp(-i \vec k \cdot \vec r).\eqno({\rm C7}) $$
Therefore, one has that 
$$H(0,0,0) = \int {\rm d}^3\vec r \; \rho_{\rm sk}(\vec r|\theta)
= M_{\rm kon},\eqno({\rm C8}) $$
where $M_{\rm kon}$ is the total mass of the semi-konus density.

The weight function we used to construct the generalized konus density is
$g(k) = k^{\mu - 1} {\rm e}^{-bk}$. Since $k_{\pm} \propto k$ we derive
from equations (C6) and (C8) that
$$M_{\rm kon} = C \cdot \lim_{k\to 0}{k^{\mu - 2} \biggl[ {\rm e}^{-bf(0)k} - 
{\rm e}^{-bg(0)k}\biggr]}.
\eqno({\rm C9})$$
where C is a constant that depends on $\mu$, $b$, and $\theta$, and
$f$ and $g$ are functions of the angle $\psi$.
Therefore, for $\mu = 1$ the total mass of the konus density is infinite.
For $\mu = 2$ one has $M_{\rm kon} > 0$, but finite, and for $\mu > 2$ the 
total mass is equal to zero.

We now show that the higher order konus densities 
$\rho_{\rm k}^{n}(r,z|\theta)$ ($n \geq 2$) of the generalized konus densities 
have zero total mass. Therefore we write the Fourier Transform of the 
semi-konus density $\rho_{\rm sk}^{n}(r,z|\theta)$ as $H_{b}^{n}(\vec k)$, 
where $n$ denotes the order of the semi-konus density, and $b$ the scalelength.
From equation (C6) it is straightforward to show that
$$H_{b}^{1}(a \vec k) = a^{\mu - 2} H_{ab}^{1}(\vec k),
\eqno({\rm C10})$$
where $a \geq 0$ is a constant, and $a \vec k = (ak, \psi, \phi)$.
The FT of the second order semi-konus density is the convolution
of $H_{b}^{1}(\vec k)$ with itself, i.e.,
$$H_{b}^{2}(\vec k) = \int {\rm d}^3\vec k' \; H_{b}^{1}(\vec k - \vec k') 
\; H_{b}^{1}(\vec k').\eqno({\rm C11})$$
Then
$$H_{b}^{2}(a \vec k) = \int {\rm d}^3\vec k' \; H_{b}^{1}(a \vec k - \vec k') 
\; H_{b}^{1}(\vec k').\eqno({\rm C12})$$
Since $\vec k'$ is a dummy variable of integration, we can substitute
$\vec k' = a \vec z$ to find
$$\eqalign{H_{b}^{2}(a \vec k) 
&= a^3 \int {\rm d}^3\vec z \; H_{b}^{1}(a (\vec k - \vec z)) 
\; H_{b}^{1}(a \vec z)\cr
&= a^{2\mu - 1} \int {\rm d}^3\vec z \; H_{ab}^{1}(\vec k - \vec z) 
\; H_{ab}^{1}(\vec z)\cr
&= a^{2\mu - 1} H_{ab}^{2}(\vec k).\cr}\eqno({\rm C13})$$
From equation (C13) we see that whereas $H_{b}^{1}(\vec k) \propto k^{\mu -1}$
at small radii, we find that $H_{b}^{2}(\vec k) \propto k^{2\mu - 1}$ at small
radii. From this analysis it is clear that the total mass $M_{\rm tot} = 
H_b^n(0,0,0)$ is equal to zero for $n \geq 2$.

\bye